\documentclass{aa}  
\usepackage{natbib}
\bibpunct{(}{)}{;}{a}{}{,} % to follow the A&A style
\usepackage{graphicx}
\usepackage{txfonts}
\usepackage{color}

\begin{document} 

\title{Investigating dust trapping in transition disks with millimeter-wave polarization}

\author{A. Pohl\inst{1,2} \and A. Kataoka\inst{2,3} \and P. Pinilla\inst{4}
	\and C. P. Dullemond\inst{2} \and Th. Henning\inst{1} \and T. Birnstiel\inst{1}}
\institute{Max-Planck-Institute for Astronomy, K\"onigstuhl 17, D-69117 Heidelberg, Germany\\
	\email{pohl@mpia.de}
    \and
		Heidelberg University, Institute of Theoretical Astrophysics, Albert-Ueberle-Str. 2, D-69120 Heidelberg, Germany
	\and
		National Astronomical Observatory of Japan, Mitaka, Tokyo 181-8588, Japan
	\and
		Leiden Observatory, Leiden University, P.O. Box 9513, NL-2300 RA Leiden, The Netherlands}

\abstract
   {Spatially resolved polarized (sub-)mm emission has been observed for example in the protoplanetary disk around HL Tau. Magnetically aligned grains are commonly interpreted as the source of polarization. However, self-scattering by large dust grains with a high enough albedo is another polarization mechanism, becoming a compelling method independent of the spectral index to constrain the dust grain size in protoplanetary disks.}
   {We study the dust polarization at mm wavelength in the dust trapping scenario proposed for transition disks, when a giant planet opens a gap in the disk. We investigate the characteristic polarization patterns and their dependence on disk inclination, dust size evolution, planet position, and observing wavelength.}
   {We combine two-dimensional hydrodynamical simulations of planet-disk interactions with self-consistent dust growth models. These size-dependent dust density distributions are used for follow-up three-dimensional radiative transfer calculations to predict the polarization degree at ALMA bands due to scattered thermal emission.}
   {Dust self-scattering has been proven to be a viable mechanism for producing polarized mm-wave radiation. We find that the polarization pattern of a disk with a planetary gap after 1\,Myr of dust evolution shows a distinctive three ring structure. Two narrow inner rings are located at the planet gap edges. A third wider ring of polarization is situated in the outer disk beyond 100\,au. For increasing observing wavelengths all three rings slightly change their position, where the innermost and outermost rings move inward. This distance is detectable comparing the results at ALMA bands 3, 6 and 7. Within the highest polarized intensity regions the polarization vectors are oriented in the azimuthal direction. For an inclined disk there is an interplay between polarization originating from a flux gradient and inclination-induced quadrupole polarization. For intermediate inclined transition disks the polarization degree is as high as $\sim 2\%$ at $\lambda=3.1\,$mm (band 3), which is well above the detection limit of future ALMA observations.}
  {}
   
%
%------------------------------------------------------------------------------------
%
\maketitle
%
%------------------------------------------------------------------------------------
%
\section{Introduction}
\label{sec:introduction}

To investigate dust coagulation and ongoing planet formation processes observational constraints on the size of grains embedded in a protoplanetary disk are crucial. Observations at millimeter (mm) wavelengths have probed large (mm sized) grains in the disk midplane, through calculation of a low spectral index of the dust opacity (e.g. \citealt{beckwith1991,testi2001,rodmann2006,ricci2010b,ricci2010,guilloteau2011}). For a continuous disk with a monotonically decreasing radial gas pressure, mm dust particles in the outer disk can experience an excessive radial drift towards the star, in contradiction to observations that revealed the existence of mm sized particles in the outer disk regions \citep{wilner2005,andrews2005,ricci2010,ricci2011}. A particle trap caused by a pressure bump might be a solution to prevent this drift problem and to allow the grains to grow efficiently \citep{whipple1972,klahr1997,fromang2005,johansen2009,pinilla2012,zhu2012}. This bump can for example result from the presence of a massive planet carving a gap in the gas density. Hence, so-called transition disks characterized by large radial gaps are excellent targets to study the impact of planet formation on the disk structure. To explain pronounced dust rings in protoplanetary disks there are certainly also other mechanisms, such as the generation of gaps at the dead-zone edges in magnetized disks without any planet \citep{flock2015}.  Particle trapping can be probed by measuring low spectral index values inside the trap, however, small opacity index values could be also explained by optically thick emission from compact regions \citep{ricci2012}. Among other uncertainties, such as the composition and porosity of dust aggregates \citep{henning1996,kataoka2014}, this shows that constraining the grain size only with opacity index measurements is still problematic.\\

Recently, \citet{kataoka2015} introduced an alternative, independent method to constrain the grain size distribution in protoplanetary disks based on dust polarization at mm wavelengths. The classical picture from the optical and near-infrared (NIR), where stellar photons are scattered by small dust in the disk surface layers, is transferred to dust self-scattering\footnote{Self-scattering means that the source of incident radiation is the thermal emission of the dust itself, which is then scattered off large dust grains resulting in polarized mm disk emission.} in the mm wavelength regime. As mentioned above, it is known that dust grains in protoplanetary disks can grow to sizes comparable to mm wavelengths, meaning that they are expected to have a large albedo and thus, can produce scattered light. When the radiation field is anisotropic, the continuum emission is expected to be partially polarized due to self-scattering of dust thermal emission. So far, polarized (sub-)mm emission has been observed in the disks around a few young stellar objects, e.g. IRAS 16293-2422B \citep{tamura1995,rao2014}, HL Tau \citep{tamura1995,stephens2014}, DG Tau \citep{tamura1999, hughes2013}, MWC 480 \citep{hughes2013}, and L1527 \citep{seguracox2015}. Commonly, magnetically aligned grains (see e.g. \citealt{lazarian2007}) are assumed as the source of polarization, but dust self-scattering is another important mechanism that needs to be considered. This idea has been explicitly applied to the protoplanetary disk around HL Tau by \citet{kataoka2016} and \citet{yang2016}, who successfully reproduced the polarization signatures observed by means of scattered mm radiation.\\

In this paper we study the dust trapping scenario when a massive planet is embedded in the disk, and investigate where polarization due to scattering can be detected. We combine 2D hydrodynamical simulations of planet-disk interactions with self-consistent dust growth models (cf. \citealt{pinilla2012b,dejuanovelar2013}). These results are used to perform 3D radiative transfer calculations in order to predict the polarization at ALMA wavelengths due to scattered thermal emission. In difference to \citet{kataoka2015}, we consider the simulated spatial dust density distribution for each grain size from the dust evolution model instead of a parametrized dust density and a simplified power law grain size distribution. We compare our results for different dust evolution timescales and analyze the dependence of the polarization degree on the disk inclination, dust composition, planet's position and observing wavelength. Moreover, it is discussed whether the polarization is detectable with future ALMA observations.\\

This paper is organized as follows. In Sect. \ref{sec:methods} we describe the numerical methods to obtain a transition disk model and the radiative transfer calculations. Section \ref{sec:results} presents our results of the dust growth modeling and the simulated polarization maps. Furthermore, we discuss our findings in terms of disk inclination, dust evolution timescales, dust composition, planet's position and observing wavelength. Finally, our results and the conclusions of this work are summarized in Sect. \ref{sec:conclusions}.

%------------------------------------------------------------------------------------

\section{Numerical methods}
\label{sec:methods}

Our numerical models start with performing hydrodynamical simulations of planet-disk interactions followed by self-consistent dust growth models, based on the setup from \citet{pinilla2012b}, also used in \citet{dejuanovelar2013}. The resulting dust distributions after different evolution times are taken as input for radiative transfer calculations with the Monte-Carlo based radiative transfer code \textsc{radmc-3d}\footnote{The code and more information is available on http://www.ita.uni-heidelberg.de/$\sim$dullemond/software/radmc-3d/.}. The details of the individual simulation steps are presented in the following Sects. \ref{subsec:methods_pd} and \ref{subsec:methods_rt}.

\subsection{Planet-disk interaction and dust evolution models}
\label{subsec:methods_pd}

The two-dimensional hydrodynamical grid code \textsc{fargo} \citep{masset2000} is used to simulate the planet-disk interaction processes. The parameters are similar to the models in \citet{pinilla2012b} and can be found in Table \ref{tab:fargo_params}. The basic model consists of a viscous disk with $\alpha=10^{-3}$ and an embedded massive planet with a planet-to-star mass ratio of $10^{-3}$ located on a fixed orbital radius at 20\,au or 60\,au, respectively. Our choice for the $\alpha$ value is motivated by recent results from \citet{dejuanovelar2016}, who constrained $\alpha$ to be within an order of magnitude of $10^{-3}$ in order to reproduce the observed structures in transition disks. The radial grid with $N_{\mathrm{r}}=512$ is logarithmically spaced between 0.5 and 140\,au, while the azimuthal grid considers $N_{\mathrm{\varphi}}=1024$ uniformly distributed cells. Furthermore, a surface density profile of $\Sigma_{\mathrm{g}} \propto r^{-1}$ is chosen, such that the disk mass corresponds to $M_{\mathrm{disk}}=0.05\,M_{\odot}$. The hydrodynamical simulations are run for about 1000 planetary orbits until the disk gas surface density has reached a quasi-stationary state.\\

In the next step, the final gas surface density is azimuthally averaged, interpolated to a radial grid of 300 cells, and used as initial condition for the 1D dust evolution code from \citet{birnstiel2010}. Since we are only interested in the radial dust distribution, this simplification is adequate. The code computes the coagulation and fragmentation of dust grains due to radial drift, turbulent mixing and gas drag. The gas surface density is kept constant during the dust evolution on simulation timescales of a few million years. This is justifiable because the gap opening timescale is much shorter compared to the dust evolution processes. As initial conditions a dust-to-gas ratio of 0.01 with an initial particle size of $1\,\mu$m and an intrinsic dust volume density of $\rho_{\mathrm{s}}=1.2\,\mathrm{g\,cm}^{-3}$ are assumed. The fragmentation velocity is considered to be $v_{\mathrm{f}}=1000\,\mathrm{cm\,s}^{-1}$. Stellar parameters typical for those of T Tauri stars ($T_{\star}=4730\,$K, $M_{\star}=1.0\,$M$_{\odot}$, $R_{\star}=1.7\,$R$_{\odot}$) are assumed. The dust grain distribution $n(r,z,a)$ depends on the grain size $a$, the distance to the star $r$ and the height above the midplane $z$. For its description the vertically integrated dust surface density distribution $\sigma_{\mathrm{d}}(r,a)$ is used, since the coagulation processes are concentrated in the disk midplane. The vertically integrated dust surface density distribution per logarithmic bin of grain radius is defined as (see \citealt{birnstiel2010})

\begin{equation}
	\sigma_{\mathrm{d}}(r,a) = \int_{-\infty}^{\infty}\,n(r,z,a)\,m\,a\,dz\,,
	\label{eq:vertintdust}
\end{equation}

\noindent with $m$ being the mass of a single particle of size $a$.\\

The dust coagulation simulations consider 180 different particle sizes from micron size up to two meters. Since the density of grains larger than 10\,cm is very small and their contribution to the emitted and scattered light is negligible, the grain size and density distributions are binned such that 140 grains with a maximum size of 7\,cm are taken into account for further calculations.

\begin{table}
	\caption{Overview of main simulation input parameters}              
	\label{tab:fargo_params}      
	\centering
	\begin{tabular}{lcc}          
		\hline\hline                       
		Parameter & Variable & Value\\
		\hline                                  
		Inner disk radius & $r_{\mathrm{in}}$ & 0.5\,au\\
		Outer disk radius & $r_{\mathrm{out}}$ & 140\,au\\
		\# of radial grid cells& $N_{\mathrm{r}}$ & 512\\
		\# of azimuthal grid cells & $N_{\mathbf{\varphi}}$ & 1024\\
		$\mathbf{\alpha}$-viscosity & $\mathbf{\alpha}$ & 10$^{-3}$\\
		Surface density at $r=r_\mathrm{p}$ & $\Sigma_{\mathrm{g,0}}$ & $1.26 \cdot 10^{-3}\,M_{\star}/r_{\mathrm{p}}^2$\\
		Surface density index & $\mathbf{\beta}$ & -1\\
		Star temperature & $T_{\star}$ & 4730\,K\\
		Star mass & $M_{\star}$ & 1.0\,$M_{\odot}$\\ 
		Star radius & $R_{\star}$ & 1.7\,$R_{\odot}$\\
		Planet-to-star mass ratio & $M_{\mathrm{p}}/M_{\star}$ & $10^{-3}$\\
		Planet position & $r_{\mathrm{p}}$ & $\lbrace 20, 60\rbrace\,\mathrm{au}$\\
		Fragmentation velocity & $v_\mathrm{f}$ & 1000\,cm/s\\
		Solid density of dust particles & $\rho_{\mathrm{s}}$ & 1.2\,g/cm$^3$\\
		\# of photons for RT & $\mathrm{nphot}_{\mathrm{scat}}$ & $5 \cdot 10^7$\\
		Observer distance & $d$ & 140\,pc\\
		Disk inclination & $i$ & $\lbrace 0, 20, 40, 60, 90\rbrace^{\circ}$\\
		\hline                                             
	\end{tabular}
\end{table}

\subsection{Radiative transfer calculations}
\label{subsec:methods_rt}

In order to calculate intensity and polarized intensity images of our transition disk model, the radiative transfer code \textsc{radmc-3d} is used. The code is frequently applied to compare theoretical disk models with observations or to make observational predictions (see recent studies from e.g. \citealt{marino2015,casassus2015,kataoka2015,juhasz2015,pohl2015}). In addition, \citet{kataoka2015} confirmed the validity of the polarization calculations of \textsc{radmc-3d} with a benchmark test proposed by \citet{pinte2009}. Intensity images and polarization maps in this paper are simulated with $5 \cdot 10^{7}$ photon packages. The linear polarization degree $P$ can be calculated with $P = \frac{\sqrt{\mathrm{Q}^2+\mathrm{U}^2}}{I}$. The radiative transfer calculations require a radiation source, which has typical stellar parameters of a T Tauri star (see Table \ref{tab:fargo_params}), where a black body radiation field is assumed for simplicity. Furthermore, the results of the dust evolution models are taken as input, i.e. the corresponding dust density for each grain size $\Sigma_{\mathrm{d}}(r)$ extracted from the distribution $\sigma_{\mathrm{d}}(r,a)$, which was previously introduced in Sect. \ref{subsec:methods_pd} and described in more detail in Sect. \ref{subsec:results_dust_distr}. The vertical density profile is assumed to be Gaussian, so that the dust volume density is given by

\begin{equation}
	\rho_{\mathrm{d}}(R,\varphi,z) = \frac{\Sigma_{\mathrm{d}}(R)}{\sqrt{2\,\pi}\,H_{\mathrm{d}}(R)}\,\exp \left( -\frac{z^2}{2\,H_{\mathrm{d}}^2(R)} \right)\,,
	\label{eq:volume_density}
\end{equation}

\noindent where $R$ and $z$ refer to spherical coordinates and can be converted into cylindrical ones via $R = r\,\sin(\theta)$ and $z = r\,\cos(\theta)$, and $\theta$ describes the polar angle. The value of the dust scale height $H_{\mathrm{d}}(R)$ is grain size dependent, since the effect of dust settling is included. Large enough dust grains are decoupled from the gas and settle down to the midplane. Following \citet{youdin2007} and \citet{birnstiel2010} the dust scale height can be estimated by

\begin{equation}
	H_{\mathrm{d}}=H_{\mathrm{p}} \cdot \mathrm{min} \left( 1,\sqrt{\frac{\alpha}{\mathrm{min}(\mathrm{St},1/2)(1+\mathrm{St}^2)}}\,, \right)  
	\label{eq:dust_scaleheight}
\end{equation}

\noindent where $\alpha$ is the viscous parameter, $H_{\mathrm{p}}$ is the gas scale height and $St$ corresponds to the Stokes number. The latter is a dimensionless coupling constant which describes how well the dust particles couple to the gas. Our study focuses on the so-called Epstein regime, where the molecular hydrogen mean free path $\lambda_{\mathrm{mfp}}$ divided by the particle size $a$ is $\lambda_{\mathrm{mfp}}/a \gtrsim \tfrac{4}{9}$. In this case the Stokes number at the midplane is given by (\citealt{birnstiel2010})
 
\begin{equation}
	\mathrm{St} = \frac{\rho_{\mathrm{s}} a}{\Sigma_{\mathrm{g}}} \frac{\pi}{2}\,,
	\label{eq:stokesnumber}
\end{equation}

\noindent where $\Sigma_{\mathrm{g}}$ denotes the gas surface density.\\

\subsubsection{Opacity calculation}
\label{subsubsec:opacity_calc}

For the opacity calculation we use Mie theory considering the BHMIE code of \citet{bohren1984} which assumes the dust grains to be spherically symmetric. The dust composition consists of a mixture between silicate (\citealt{draine2003b}), carbonaceous material (\citealt{zubko1996}) and water ice (\citealt{warren2008}), consistent with \citet{ricci2010}, who considered the bulk densities from \citet{pollack1994}. The fractional abundances are taken as 7\%, 21\% and 42\% so that the amount of vacuum is 30\%. The opacity of the mixture is determined by means of the Bruggeman mixing theory. The absorption and scattering opacities, $\kappa_{\mathrm{scat}}$ and $\kappa_{\mathrm{abs}}$, as well as the scattering matrix elements $Z_{ij}$ are calculated. Here, the notation $Z_{ij}$ of the \textsc{radmc-3d} code is used, which is related to the notation $S_{ij}$ from \citet{bohren1984} by $Z_{ij}=S_{ij}/(k^2m_{\mathrm{grain}})$ , with $k$ being the wave number. Taking the angular integral of the $Z_{11}$ matrix element gives the scattering opacity,

\begin{equation}
	\kappa_{\mathrm{scat}}(a) = \oint Z_{11}(a,\theta)\,d\Omega\ = 2\pi \int_{-1}^{1} Z_{11}(\mu)\,d\mu\,,
	\label{eq:kscat}
\end{equation}

\noindent where $\mu = cos(\theta)$. Similarly, the anisotropy factor $g$ is given by the following integration 

\begin{equation}
	g = \frac{2\pi}{\kappa_{\mathrm{scat}}} \int_{-1}^{1} Z_{11}(\mu)\,\mu\,d\mu\,.
	\label{eq:gfactor}
\end{equation}

As already mentioned in Sect. \ref{subsec:methods_pd}, 140 different grain sizes are considered. For the scattering phase function represented by $Z_{11}$ as a function of scattering angle Mie theory produces very strong oscillations if the wavelength of the incoming radiation is very small compared to the grain size. Therefore, a grain size distribution between two neighboring grain size bins is applied with a MRN powerlaw index of -3.5 \citep{mathis1977}. This averaging smears out the wiggles, giving more manageable opacity values. Furthermore, to avoid problems with the spatial resolution of our models, we define a cut-off value for $Z_{11}$ setting an upper limit for the forward peaking of the phase function. This means that for very small scattering angles $Z_{11}$ is set to a constant (typically to the value it has at 5\,deg). This is reasonable since such low scattering angles are not relevant for our analysis of mostly highly inclined disks.\\

The polarization by dust scattering is strongly dependent on the scattering angle, grain size and wavelength. Therefore, in order to determine an effective polarization degree $(-Z_{12}(\theta)/Z_{11}(\theta))_{\mathrm{eff}}$, effective scattering matrix elements $Z_{11,\mathrm{eff}}(\theta)$ and $Z_{12,\mathrm{eff}}(\theta)$ have to be derived. In our case this is done by weighting the individual elements $Z_{11}(a,\theta)$ and $Z_{12}(a,\theta)$ by the corresponding dust density distribution $\Sigma_{\mathrm{d}}(a,r_0)$ at a specific radial position $r_0$. More precisely, we obtain

\begin{equation}
	Z_{11(12),\mathrm{eff}}(\theta) = \frac{\int_{a_{\mathrm{min}}}^{a_{\mathrm{max}}} \sigma_{\mathrm{d}}(a,r_0)\,Z_{11(12)}(a,\theta)\,da}{\int_{a_{\mathrm{min}}}^{a_{\mathrm{max}}} \sigma_{\mathrm{d}}(a,r_0)\,da}	\,,
	\label{eq:z11eff}
\end{equation}

\noindent so that the effective polarization degree is determined by

\begin{equation}
	(-Z_{12}(\theta)/Z_{11}(\theta))_{\mathrm{eff}} = - \frac{\int_{a_{\mathrm{min}}}^{a_{\mathrm{max}}} \sigma_{\mathrm{d}}(a,r_0)\,Z_{12}(a,\theta)\,da}{\int_{a_{\mathrm{min}}}^{a_{\mathrm{max}}} \sigma_{\mathrm{d}}(a,r_0)\,Z_{11}(a,\theta)\,da}\,.
	\label{eq:z11z12eff}
\end{equation}

\begin{figure}
	\centering
	\includegraphics[width=0.5\textwidth]{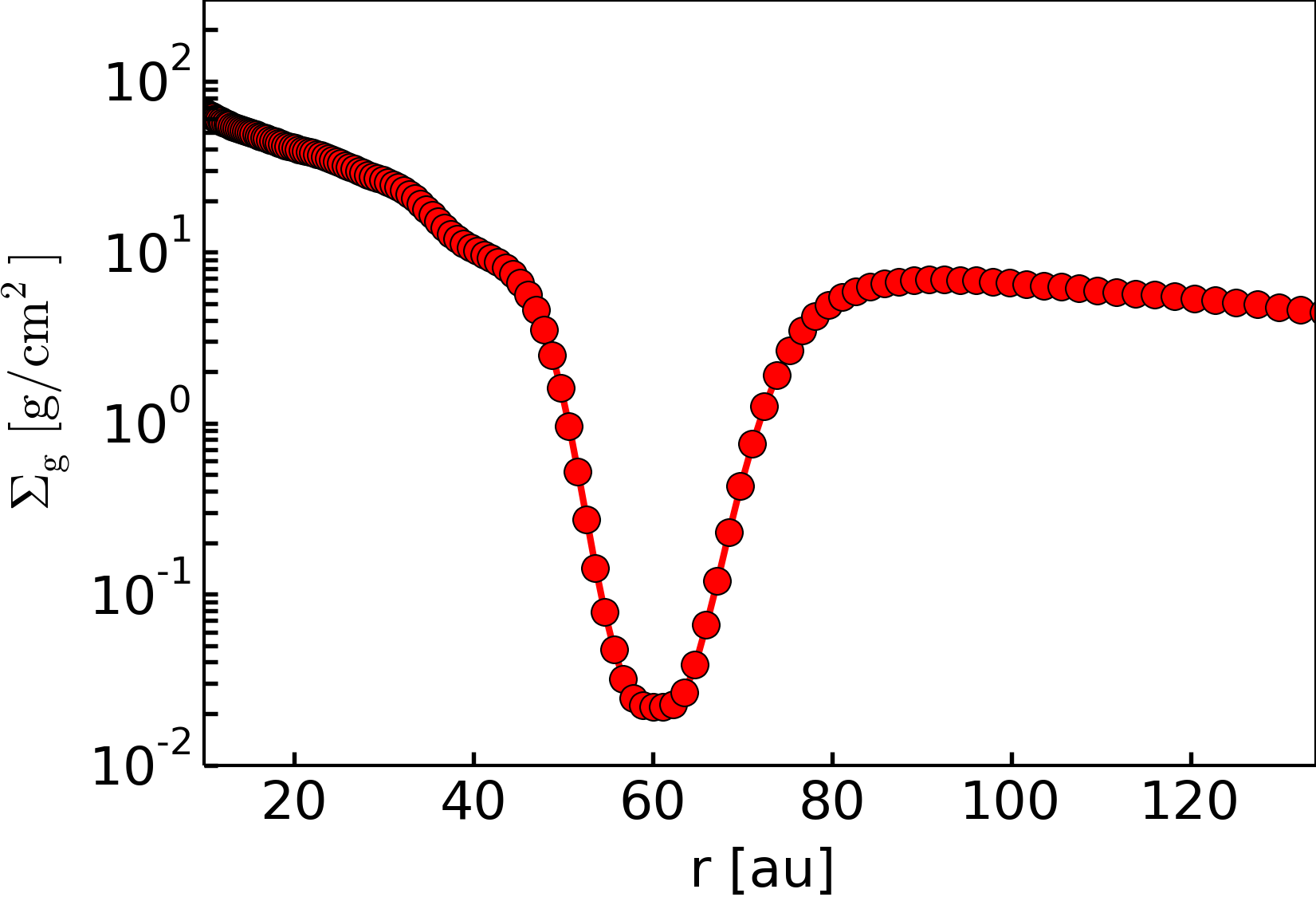}
	\caption{Azimuthally averaged radial gas surface density profile after 1000 orbits of evolution for a planet with a planet-to-star mass ratio of $10^{-3}$ located at 60\,au in a turbulent, flared disk.}
	\label{fig:gasdensity}
\end{figure}

\begin{figure*}
	\centering
	\includegraphics[width=1.0\textwidth]{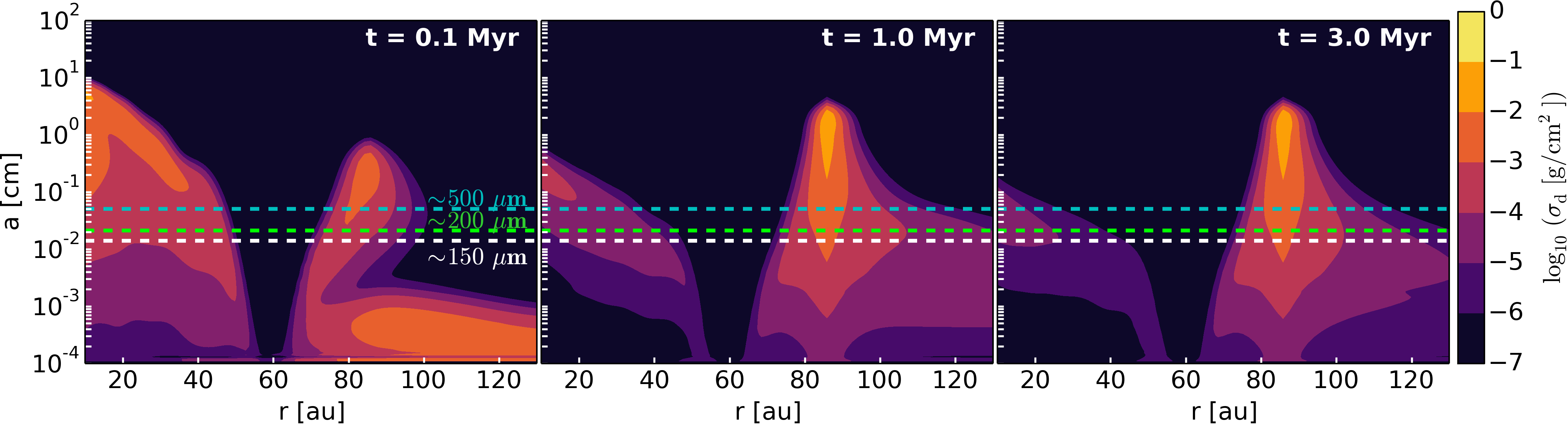}
	\caption{Snapshots of the vertically integrated dust density distribution for evolutionary times of 0.1, 1 and 3\,Myr. The planet is located at 60\,au and the planet-to-star mass ratio is $10^{-3}$. The horizontal dashed lines represent the particle sizes corresponding to a size parameter of $x=2\pi a/ \lambda \sim 1$ considering $\lambda=0.87\,$mm, $\lambda=1.3\,$mm and $\lambda=3.1\,$mm. If the dominant grain size at a certain disk location is less than $\sim \lambda/2\pi$, there will be significant polarization in scattered light.}
	\label{fig:dustdensity}
\end{figure*}

\subsubsection{Temperature profile}
\label{subsubsec:temperature}

Since we are only interested in the scattering, for simplicity we assume that the dust temperatures are the same for all grain species. Thus, for the dust temperature profile of the disk a power-law distribution is adopted, 

\begin{equation}
	T(r) = T_{0}\,\left( \frac{r}{r_0} \right)^{2f-1}\,,
	\label{eq:tempdistr}
\end{equation}

\noindent where $T_0$ is set to 65\,K, $r_0$ is equal to 20\,au and the flaring index $f$ corresponds to 0.25. The parameters are chosen such that they coincide with those of the hydrodynamical FARGO simulations and are in agreement with the disk geometry found in observations of T Tauri disks (e.g. \citealt{dalessio2001}).

%------------------------------------------------------------------------------------

\section{Results}
\label{sec:results}

\subsection{Gas and dust density distribution}
\label{subsec:results_dust_distr}

\begin{figure}
	\centering
	\includegraphics[width=0.5\textwidth]{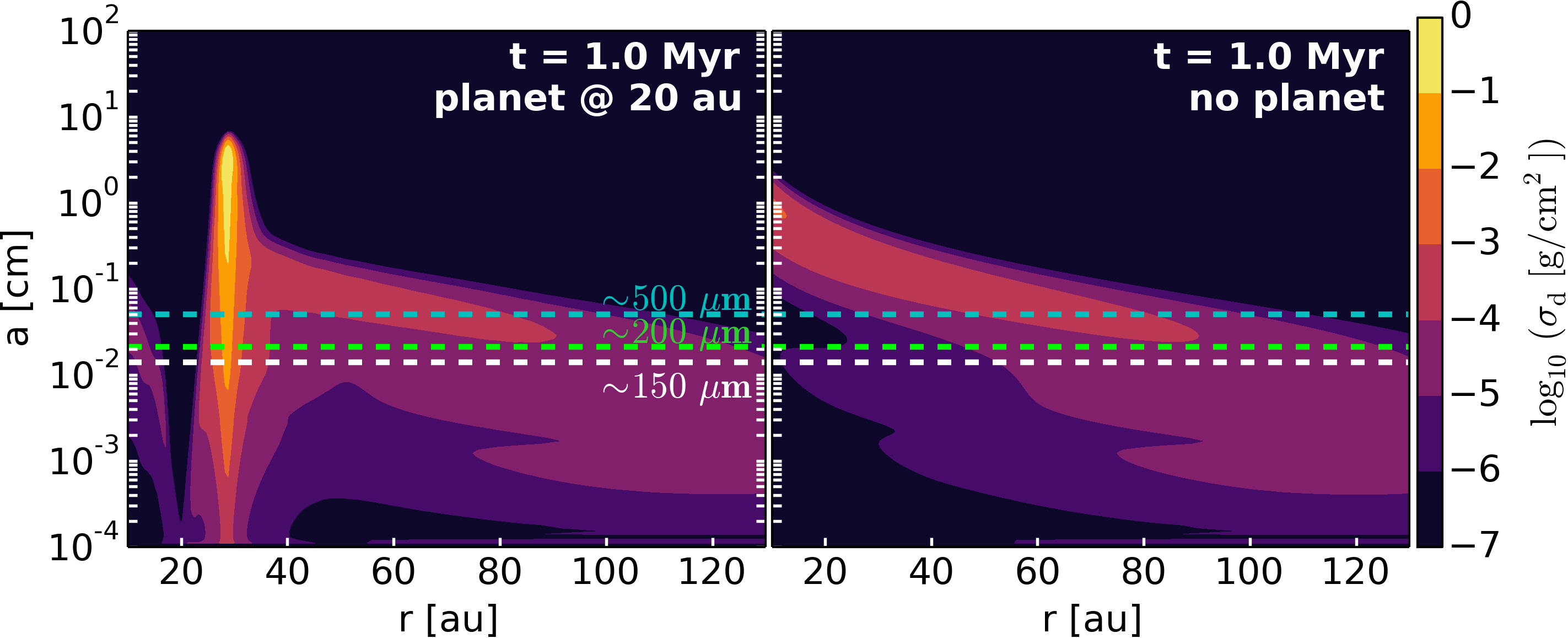}
	\caption{Vertically integrated dust density distribution after 1.0\,Myr of dust evolution for the planet embedded at 20\,au (left) and the comparison case without any planet (right).}
	\label{fig:dustdensity_np}
\end{figure}

Figure \ref{fig:gasdensity} illustrates the azimuthally averaged gas surface density profile after 1000 orbits of evolution of the planet-disk interaction processes. The embedded planet with a planet-to-star mass ratio of $10^{-3}$ (1\,M$_{\mathrm{jup}}$ for our case) is located on a fixed circular orbit at 60\,au in a flared, viscous ($\alpha=10^{-3}$) disk. A pronounced gap is opened by the planet, whereas its width is constrained by the planet mass. The presence of the planet generates a pressure bump in the otherwise monotonically decreasing pressure distribution (see \citealt{pinilla2012b} for detailed discussion). The vertically integrated dust density distribution (cf. Eq. \ref{eq:vertintdust}) for three different times of dust evolution (0.1, 1.0 and 3.0\,Myr) is shown in Fig. \ref{fig:dustdensity}. As we noticed for the gas density, the planet also clearly carves a gap in the dust influencing the distribution of dust particle sizes in the inner and outer disk. Large dust grains of mm-size are trapped at the pressure maximum, which is located at a larger radius than the gap edge in the gas density. Smaller, micron-sized particles can pass the pressure maximum and radially drift to the inner disk region. Dust particles in the outer regions of the disk grow until they reach a size of mm and remain there several million years. Additionally, the result of a comparison simulation without any planet after 1\,Myr of dust evolution in the disk is shown in Fig. \ref{fig:dustdensity_np}. In this case no pressure bump is created, instead the dust just grows, fragments and drifts towards the star without building any accumulation of mm-sized grains in the outer disk regions.

\subsection{Analysis of polarization maps}
\label{subsec:results_pol}

\begin{figure*}
	\centering
	\includegraphics[width=1.0\textwidth]{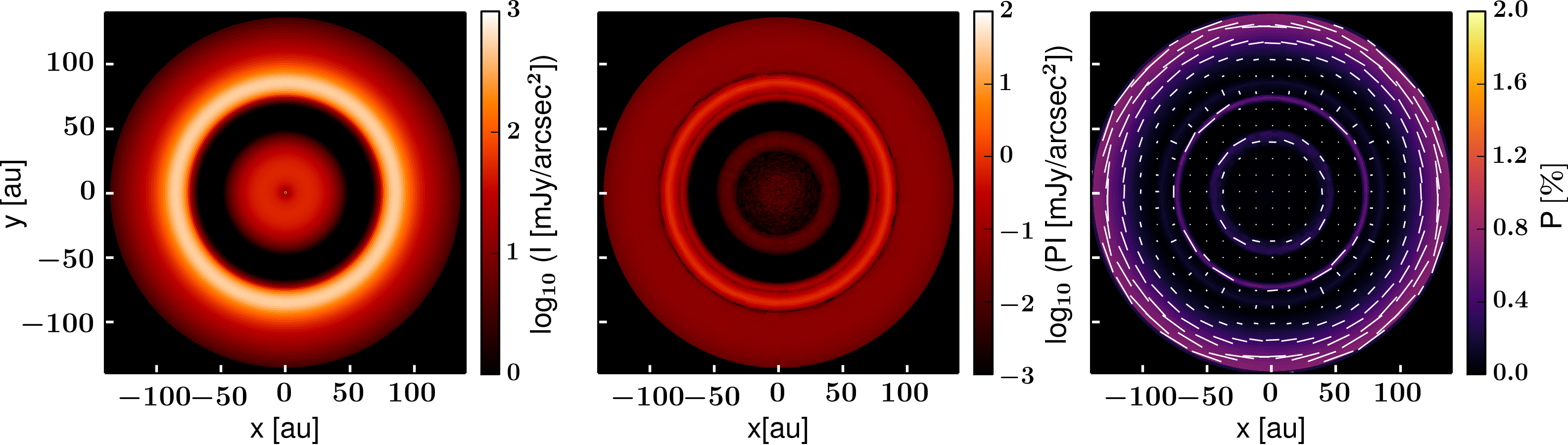}
	\caption{Intensity $I$, polarized intensity $PI$, and polarization degree $P$ overlaid with polarization vectors for the reference model (planet-to-star mass ratio of $10^{-3}$, planet located at 60\,au, 1\,Myr of dust evolution, dust opacity mixture, $\lambda=0.87\,$mm, face-on disk).}
	\label{fig:ref_intens}
\end{figure*}

As shown in \citet{kataoka2015} even without a specific central light source, the continuum emission is polarized due to self-scattering if the dust emission exhibits quadrupole anisotropy at certain positions in the disk. In this section we investigate the characteristic polarization pattern for a transition disk and discuss the detection possibility with current (sub-)mm observations. The basic polarization mechanism is explained in Sect. \ref{subsubsec:results_reference} by means of a reference face-on disk model. The effects on the polarization of disk inclination, observing wavelength, dust size evolution, dust composition and presence of the planet are studied in Sects. \ref{subsubsec:results_inclin}--\ref{subsubsec:results_planetpos}.

\subsubsection{Reference simulation}
\label{subsubsec:results_reference}

As a reference simulation we define the case of a planet with planet-to-star mass ratio of $10^{-3}$ located at 60\,au taking the dust evolution of the disk after 1\,Myr and an opacity mixture as described in Sect. \ref{subsubsec:opacity_calc}. By means of \textsc{radmc-3d} we calculate intensity and polarized intensity images at a reference wavelength of $\lambda=0.87\,$mm (ALMA band 7) considering anisotropic scattering with polarization. Figure \ref{fig:ref_intens} displays the Stokes $I$ intensity, polarized intensity $\mathrm{PI}=\sqrt{Q^2+U^2}$, and polarization degree $P=PI/I$ with overlaying polarization vectors for a face-on disk with respect to the line-of-sight. The intensity image reveals the structure of the transition disk, such that around the planet's position a gap is detected. As explained in the previous Sect. \ref{subsec:results_dust_distr}, mm grains are trapped in the pressure maximum at $\sim 85$\,au and generate the bright emission ring at that location.\\

\begin{figure}
	\centering
	\includegraphics[width=0.5\textwidth]{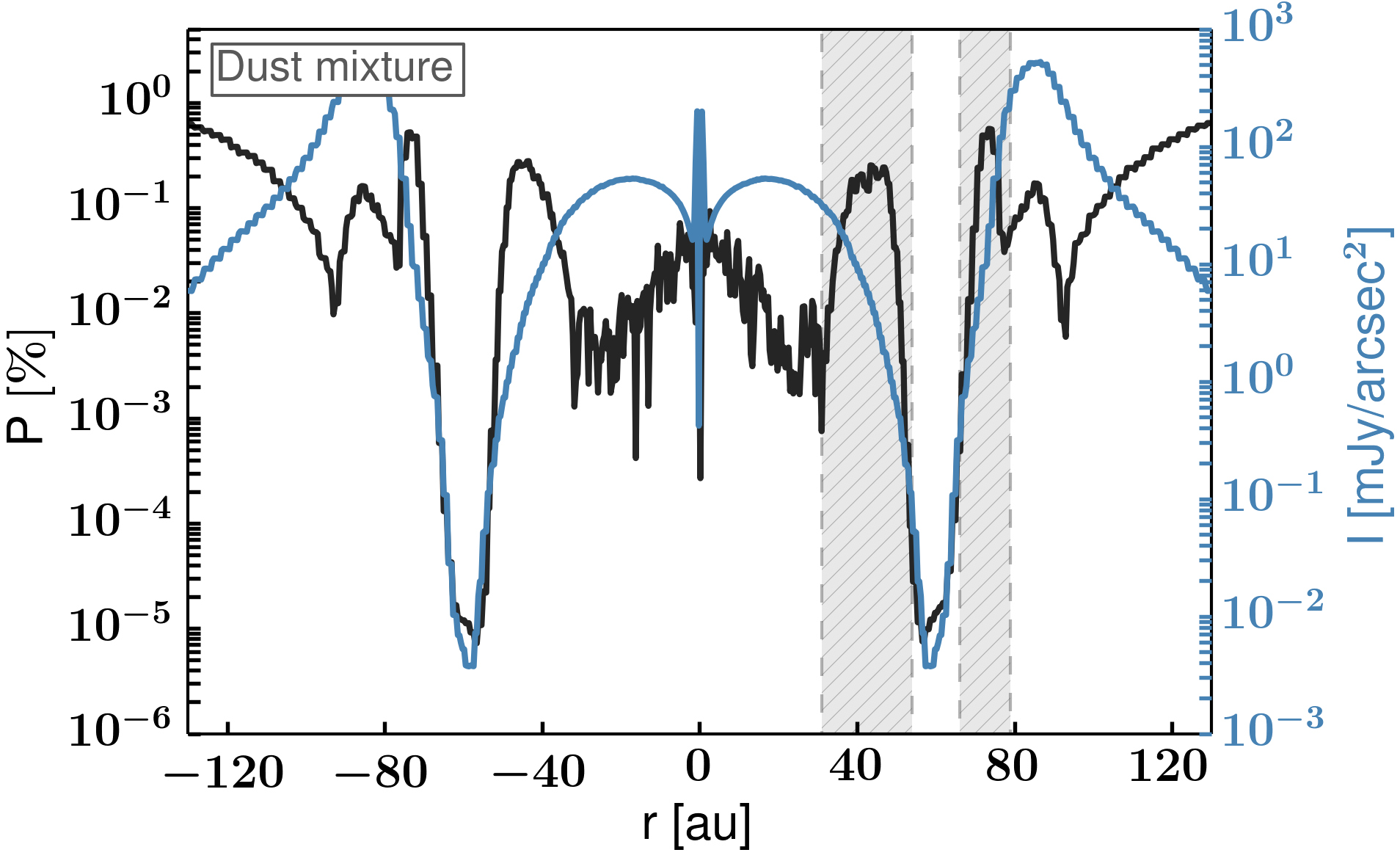}
	\caption{Radial cut of the intensity profile (blue) and polarization degree (black) along the major axis for the case in Fig. \ref{fig:ref_intens}. The gray crosshatched areas visualize the two inner polarization peaks.}
	\label{fig:ref_radcut}
\end{figure}

\begin{figure}
	\centering
	\includegraphics[width=0.5\textwidth]{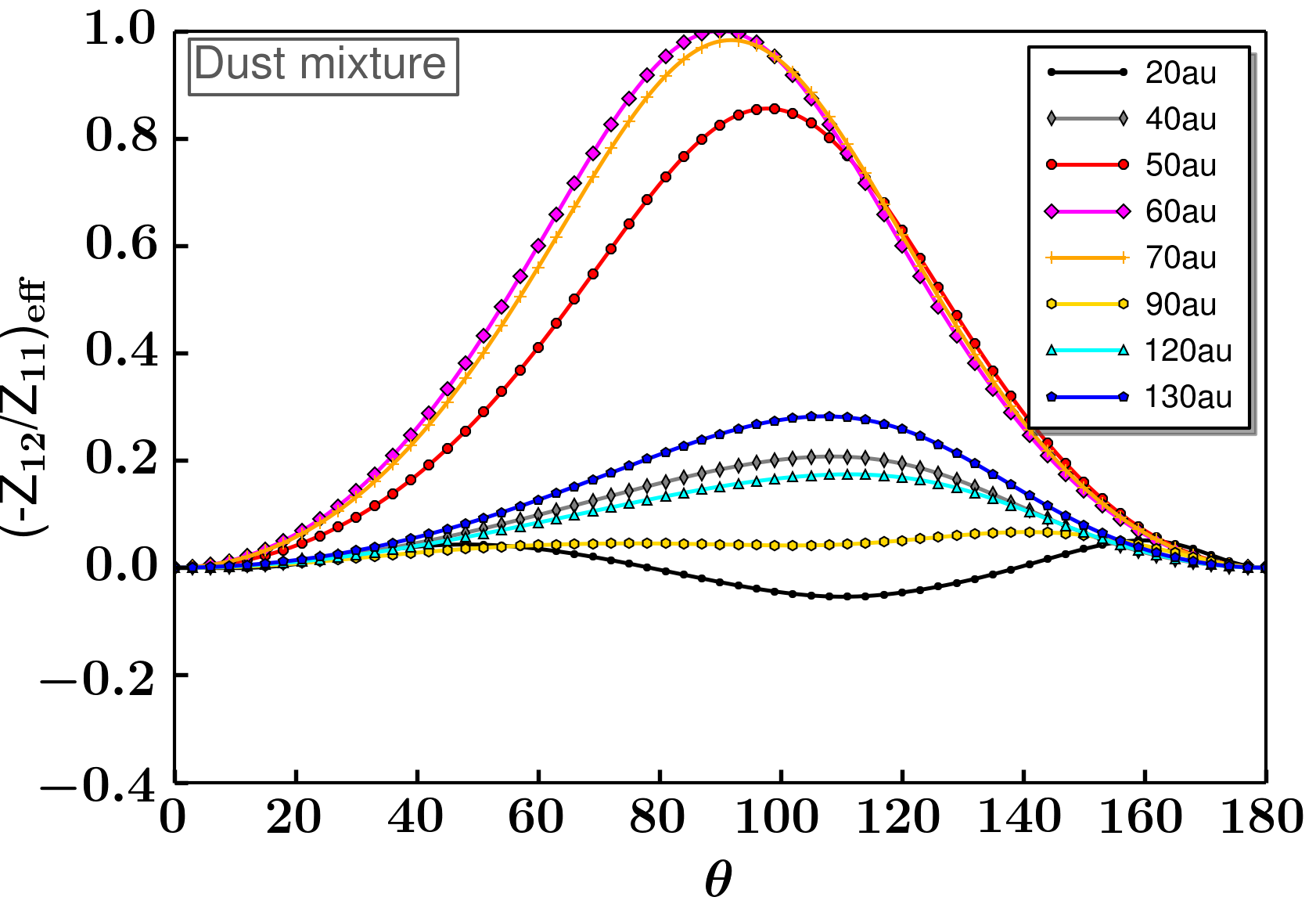}
	\caption{Effective degree of polarization $(-Z_{12}(\theta)/Z_{11}(\theta))_{\mathrm{eff}}$ of the dust grains in dependence of the scattering angle $\theta$. Several locations throughout the disk are chosen and marked in different colors.}
	\label{fig:ref_z1211}
\end{figure}

\begin{figure*}
	\centering
	\includegraphics[width=1.0\textwidth]{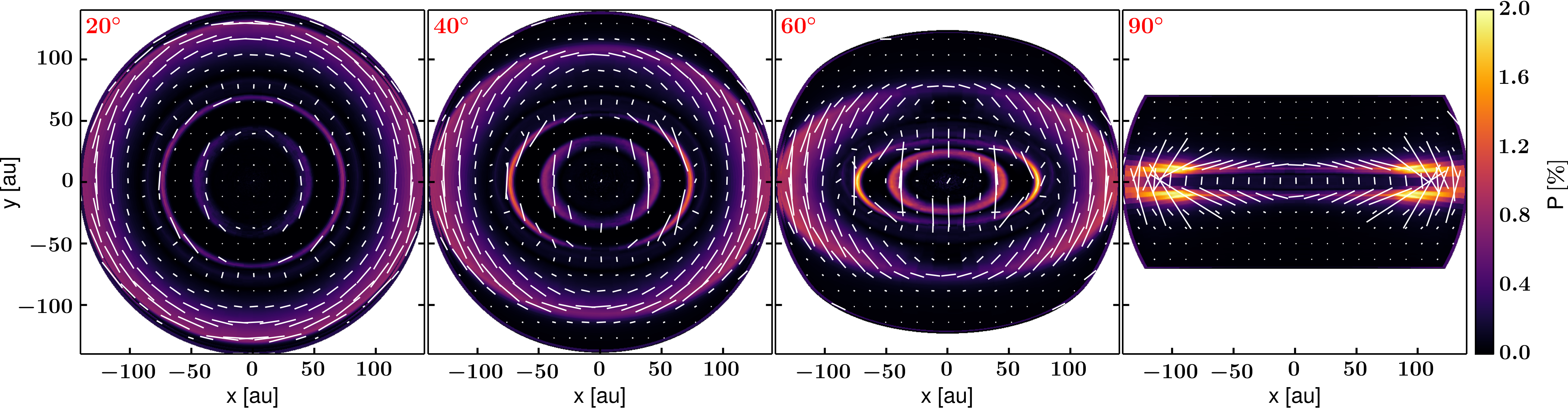}
	\caption{Polarization degree maps with overplotted polarization vectors for different disk inclination angles (20$^{\circ}$, 40$^{\circ}$, 60$^{\circ}$ and 90$^{\circ}$ from left to right) after 1\,Myr of dust evolution. The images are produced at $0.87\,$mm. The reference face-on model can be found in Fig. \ref{fig:ref_intens}.}
	\label{fig:inclin_pol}
\end{figure*}

Polarization generally depends on the observing wavelength, grain size and scattering angle. For small, micrometer-sized particles the scattering opacity is much smaller than the absorption opacity in the mm. Scattering at mm wavelength becomes efficient if the dust grains are as large as a few hundred micrometers. In addition, there is also an upper grain size limit for significant polarization due to scattering. As discussed in \citet{kataoka2015} the maximum grain size that is responsible for producing polarization is determined by $a_{\mathrm{max}} \sim \lambda/2\pi$. This is illustrated by the horizontal dashed lines in Figs. \ref{fig:dustdensity} and \ref{fig:dustdensity_np}. If grains larger than that size are present at a certain location in the disk, no polarization is expected. To distinguish between different scattering regimes, the size parameter $x = 2\pi a/\lambda$ is important (\citealt{bohren1984}). For the Rayleigh scattering regime with $x < 1$, i.e. for the grain size being smaller than the wavelength, the polarization curve as a function of scattering angle $(-Z_{12}(\theta)/Z_{11}(\theta))_{\mathrm{eff}}$ is symmetric and features a peak at $\theta = 90^{\circ}$. This can be clearly seen in Fig. \ref{fig:ref_z1211} for most disk locations for a wavelength of 0.87\,mm. One also recognizes that for certain radii where mm/cm grains dominate, e.g. at the pressure bump region around 90\,au (yellow line) and in the very inner disk at 20\,au (black line), there is a sign flip in the polarization, which is known as polarization reversal (\citealt{daniel1980,fischer1994,kirchschlager2014}). This generally influences the orientation of the polarization vectors. However, effects of negative polarization are not expected in our simulations since the absolute polarization values are quite small at those positions in the disk where negative polarization occurs.\\

The polarization map in the right panel of Fig. \ref{fig:ref_intens} shows a three ring structure. These features are unique for the polarization due to dust self-scattering in a transition disk hosting a gap and pressure bump caused by a giant planet. The rings' positions coincide with those locations of the disk where the dust grains have a maximum size of $a_{\mathrm{max}} \approx 150\,\mu$m for $\lambda = 0.87\,$mm. This can be verified by looking at the dashed blue line in the center panel of Fig. \ref{fig:dustdensity} and counting the crossing points/regions for which this condition is fulfilled. This is the case just around the gap and in the outer disk beyond the pressure bump, illustrated also in Fig. \ref{fig:ref_radcut}. The third ring is wider in the radial dimension because the suitable maximum grain size is met for the whole outer part $\gtrsim\,110\,$au. We note that the high polarization degree is unaffected by the disk boundaries, proven by a model with an extended disk radius of 300\,au instead of 140\,au. At the location of the pressure bump big cm and mm grains ($\gg 150\,\mu$m), which do not produce polarization, are present as well. Although the polarized intensity is not zero, since there are particles with a suitable size for polarization, the amount of unpolarized intensity is high. Hence, the ratio of polarized to unpolarized radiation is tiny and no polarization is expected in that region. Another important prerequisite to have such a significantly polarization degree is that the radiation field has a strong gradient field. Thus, the third ring has the highest polarization degree since it is located outside of the bright intensity ring. The dominant photon source for the outer disk is determined by the thermal emission from this intensity peak.\\

The two inner rings are inside of the bright emission ring. For the innermost ring, it is expected that the incident radiation originates primarily from the very inner disk acting as a point source. The polarization vectors in all rings are orientated in azimuthal direction for this face-on disk. Two effects play an important role here, the propagation direction of incident photons and the sign of the polarization. Most of the incident photons for the last scattering come from the inner disk or emission from the ring region and therefore, propagate radially outwards. Thus, assuming that we are in the Rayleigh scattering regime and hence there is no polarization reversal, the radiation is azimuthally polarized.\\

It should be emphasized that if the amount of polarized intensity is too low, no polarization can be detected, even if the polarization degree itself were high. Since the polarized intensity is highest around 85\,au, the second and third polarization ring might be observationally most important. The width of the bright emission ring in PI is $\sim 30\,$au, which defines the minimum spatial resolution needed to resolve the ring structure.

\subsubsection{Effect of disk inclination}
\label{subsubsec:results_inclin}

\begin{figure}
	\centering
	\centerline{
		\includegraphics[width=0.35\textwidth]{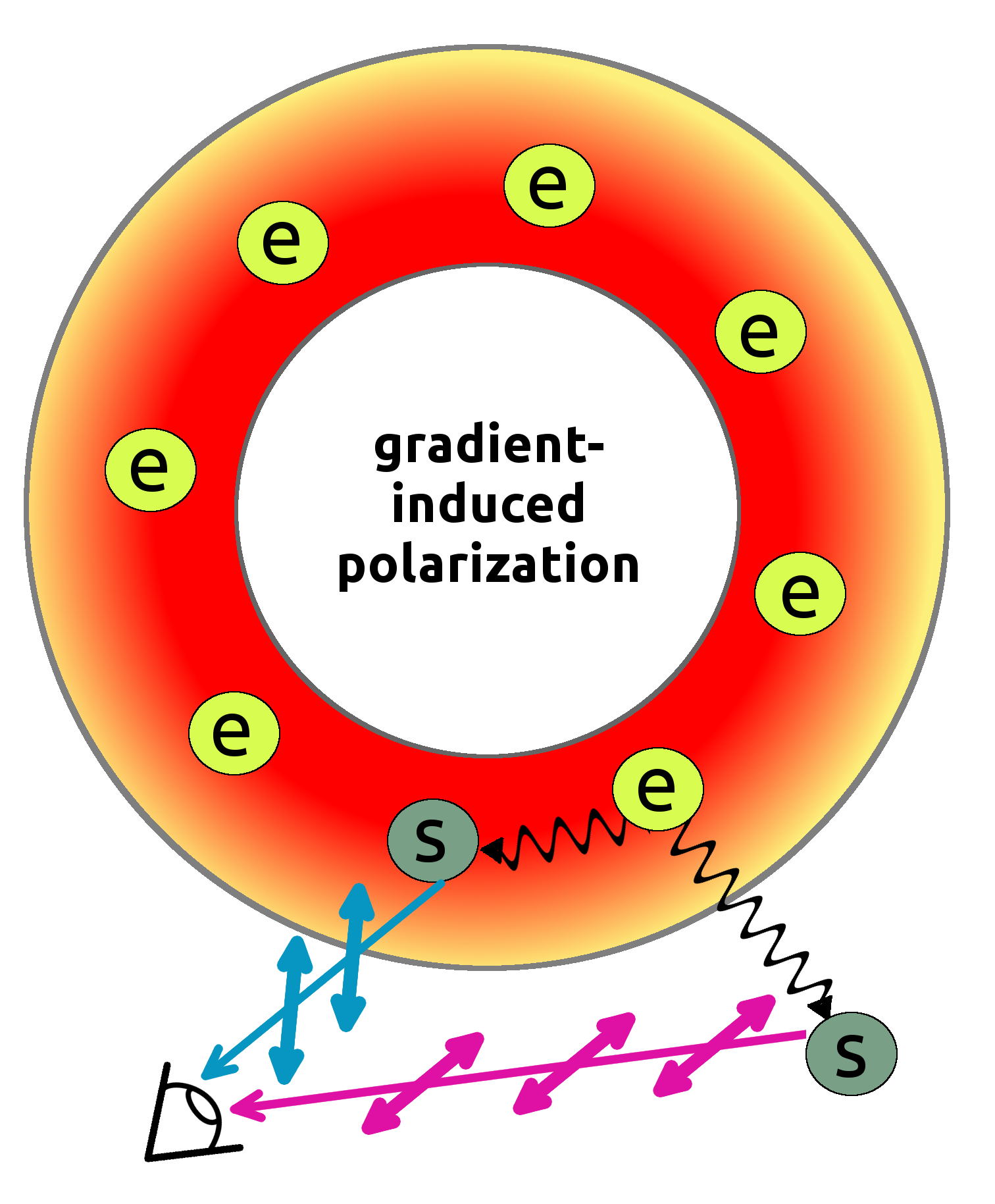}
	}
	\centerline{
		\includegraphics[width=0.35\textwidth]{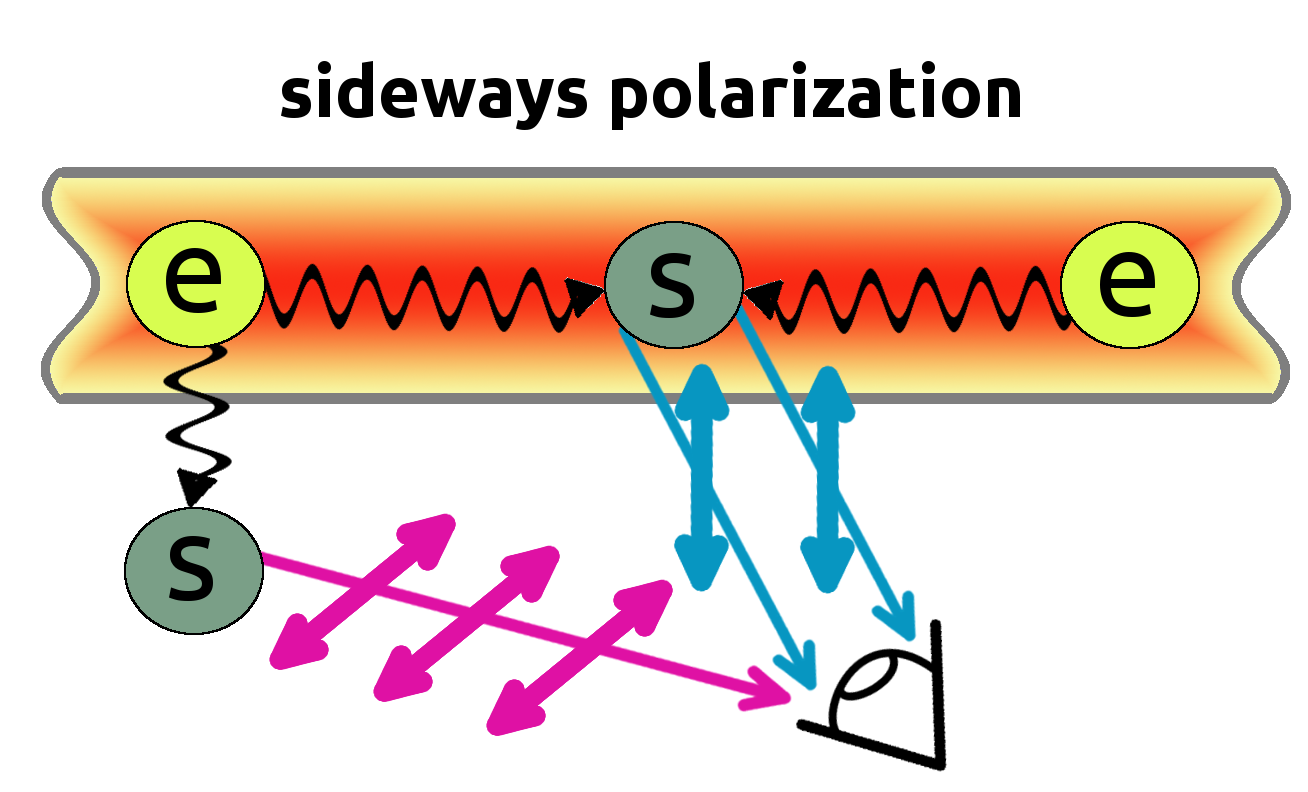}
	}
	\caption{Sketch illustrating the difference between gradient-induced polarization (top) and sideways polarization (bottom). The upper sketch considers a face-on disk, while the observer has an edge-on view for the lower image. Circles with character \textit{e} stand for dust grains working as an emitter, those with \textit{s} work as a scatterer.}
	\label{fig:sketch}
\end{figure}

Figure \ref{fig:inclin_pol} shows polarization maps for different disk inclinations (20$^{\circ}$, 40$^{\circ}$, 60$^{\circ}$ and 90$^{\circ}$) with respect to the observer. All other simulation parameters are the same as in the reference model presented in Sect. \ref{subsubsec:results_reference} and displayed in Fig. \ref{fig:ref_intens}. The general polarization degree structure, i.e. the number and position of the rings, stays the same, since the dust density distribution has not been changed. Nonetheless, the amount of polarization and the orientation of the polarization vectors drastically changes. In the previous Sect. \ref{subsubsec:results_reference} we show that a high polarization degree for a face-on disk is linked to a strong flux gradient, which causes the polarization vectors to be azimuthally orientated. In the following this mechanism is referred to as \textit{gradient-induced polarization}. For an inclined disk there is an interplay between this effect and inclination-induced quadrupole polarization, in the following called \textit{sideways polarization}. Both polarization mechanisms are sketched in Fig. \ref{fig:sketch}.\\

In an inclined disk the light coming from a direction along the major axis is scattered by 90$^{\circ}$ and therefore fully polarized along the minor axis in the Rayleigh limit. Contrarily, incident light from the minor axis is scattered by 90$^{\circ} \pm i$, which is then partially polarized along the major axis of the disk. Thus, the majority of the polarization vectors are orientated along the minor axis. Qualitatively, \citet{yang2016} found a similar polarization vector behavior for the specific case of HL Tau using a semi-analytic model. To better understand the sideways polarization, one should first take a look at the case of an edge-on disk (right most panel of Fig. \ref{fig:inclin_pol}). The basic mechanism is illustrated in Fig. \ref{fig:sketch}. The polarization effect seen in this case is comparable to that for a tube-like density distribution (cf. Fig. 5 in \citealt{kataoka2015}), just imagine to bend the tube forming a ring. The radiation inside the tube is polarized radially and so it is detected for the edge-on disk due to the net flux from the azimuthal direction being larger than that from the radial direction. Since the midplane is, however, optically thick, the polarization degree values themselves are quite low. The decrease of polarization degree at optically thick regions was also reported in \citet{kataoka2015}. Furthermore, the vector orientation changes for the location where the vertical optical depth reaches unity. Therefore, the vertical gradient of radiation dominates and the polarization vectors are azimuthally orientated. Intermediate inclined disks, such as $i=20^{\circ}$, 40$^{\circ}$ or 60$^{\circ}$, indicate that for all rings there is a decrease of polarization along the minor axis and enhanced polarization along the major axis. In the outer ring the gradient-induced polarization and, therefore, the azimuthal vector orientation still dominates, since the major emission comes from the inside. As already mentioned, the bright intensity ring works as the thermal emission source here. In contrast, the inclination-induced sideways effect is strong for the central region inside of the intensity ring, where the gradient-induced polarization is weak already for the face-on disk. Hence, there is a polarization direction change by 90$^{\circ}$ along the minor axis in the inner disk. Indeed, the fraction of scattered radiation polarized due to disk inclination gets larger with increasing inclination angle.\\

\begin{figure}
	\centering
	\includegraphics[width=0.5\textwidth]{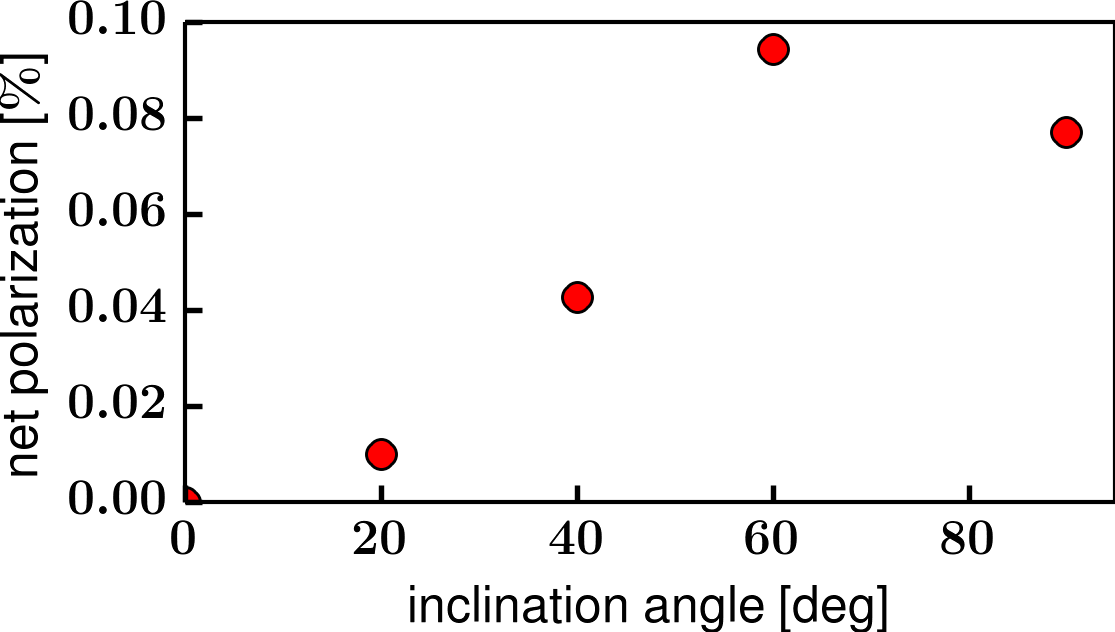}
	\caption{Net polarization dependent on the disk inclination angle for the cases shown in Fig. \ref{fig:inclin_pol}.}
	\label{fig:netpol_incl}
\end{figure}

The polarization detection generally depends on the spatial resolution. If the disk cannot be spatially resolved, i.e. it is seen as a point source, it depends on the disk inclination whether polarization can be actually detected. Considering this case, for a face-on disk, where exclusively gradient-induced polarization acts, the polarization components cancel out ending in no polarization at all. However, for an inclined disk, where the sideways polarization plays an important role, polarization can be detected even for a point source. In Fig. \ref{fig:netpol_incl} the net polarization $\sqrt{(\Sigma Q)^2+(\Sigma U)^2}/\Sigma I$ is plotted versus the disk inclination angle. It is shown that the net polarization is zero for the face-on disk, gradually increases with inclination and peaks at $60^{\circ}$. Then, it slightly decreases again for the edge-on disk. This clearly indicates that the sideways polarization is the dominant factor for the net polarization. Since we are interested in detecting the characteristic polarization features when dust tapping is triggered by a planet embedded in the disk, we do not consider edge-on disks in the following simulations. We set the disk inclination angle to an intermediate value of $40^{\circ}$ to obtain polarization at a high level and still resolve the structures. Since the polarization degree is up to $\sim 2\%$ in this case, it is likely to be detected with ALMA observations as in the case of HL Tau \citep{stephens2014,kataoka2015,yang2016}. A spatial resolution as high as 0\farcs2 is required to safely resolve the emission rings in PI, and hence the polarization structures. Due to the non-detection of polarization by previous observational studies, we strongly argue here that a high enough spatial resolution is the key for our science goals.

\subsubsection{Wavelength dependence of polarization rings}
\label{subsubsec:results_wavelength}

\begin{figure}
	\centering
	\includegraphics[width=0.5\textwidth]{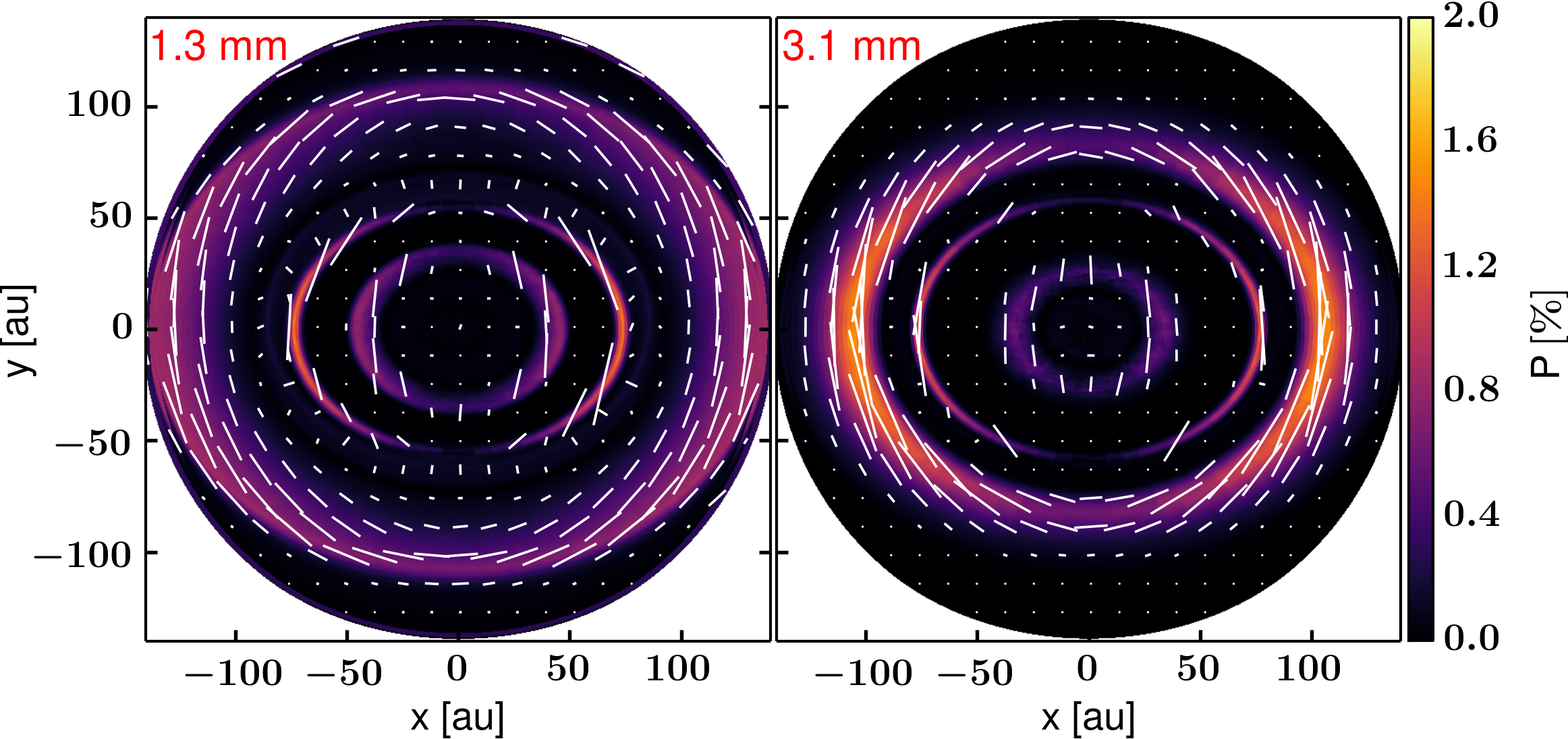}
	\caption{Polarization maps with overplotted polarization vectors at $\lambda=1.3\,$mm (ALMA band 6) and $\lambda=3.1\,$mm (ALMA band 3) after 1\,Myr of dust evolution and for a disk inclination angle of 40$^{\circ}$.}
	\label{fig:lambda_pol}
\end{figure}

Observations at different wavelengths are most sensitive to different particle sizes. Furthermore, as described in Sect. \ref{subsubsec:results_reference} the maximum grain size for producing polarization is proportional to the observing wavelength. Hence, although the basic polarization ring structure stays the same, the ring locations slightly change with the wavelength. More precisely, the innermost and outermost rings move inwards, while the middle ring moves in opposite direction for longer wavelengths as displayed in Fig. \ref{fig:lambda_pol}. For the third ring the moving distance of the polarization peak is $\sim 20\,$au, which is large enough to be detectable with the two ALMA bands 3 and 7. This is what we expect from the dust density distribution in Fig. \ref{fig:dustdensity}, center bottom panel. With our dust transport for each grain size together with coagulation and fragmentation of grains, a spatial segregation of the grain size distribution is modeled. The crossing point of the most upper horizontal dashed line (representative for $\lambda=3.1\,$mm) with the maximum available grain size is further inside than for the two lower horizontal lines (representing the shorter observing wavelengths). The maximum polarization degree is approximately by a factor of two higher for $\lambda=3.1\,$mm compared to $\lambda=0.87\,$mm. The reason for this is that the albedo $\eta = \kappa_{\mathrm{scat}}/(\kappa_{\mathrm{abs}}+\kappa_{\mathrm{scat}})$ increases with grain size and the amount of polarized intensity with respect to the unpolarized one is higher for longer wavelengths.

\subsubsection{Effect of dust size evolution}
\label{subsubsec:results_sizeevol}

\begin{figure*}
	\centering
	\includegraphics[width=1.0\textwidth]{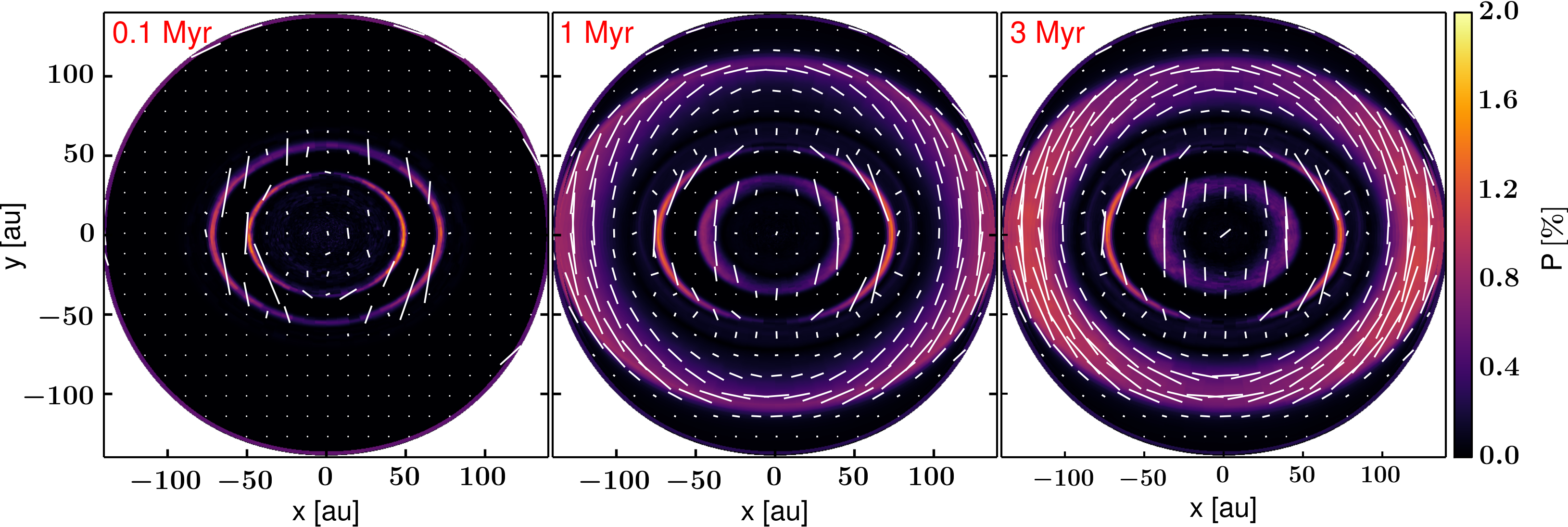}
	\caption{Polarization maps with overplotted polarization vectors after 0.1\,Myr, 1\,Myr and 3\,Myr of dust evolution and for a disk inclination angle of 40$^{\circ}$. The images are produced at 0.87\,mm.}
	\label{fig:time_pol}
\end{figure*}

To understand whether and how the polarization map changes with the dust evolution timescale we study the polarization degree at different time snapshots. We show the results for 0.1\,Myr, 1\,Myr and 3\,Myr in Fig. \ref{fig:time_pol}. At a very early dust evolution time only the two inner polarization rings are produced. The reason for the absence of the third ring is that grains have not grown to sizes larger than 10 microns in the outer disk at that stage of evolution. For the small grains the absorption opacity dominates over the scattering opacity so that their scattering at mm wavelength is inefficient. At longer times of evolution, grains grow to larger sizes and drift to the region of high pressure. The third polarization ring appears already after approximately 0.5\,Myr due to this efficient grain growth mechanism. From 1\,Myr of dust evolution on there is a quite stable situation for the overall polarization structure. At $t=3\,$Myr only the polarization degree in the innermost ring is slightly reduced due to the ongoing drift of particles towards the star. Contrarily, the maximum polarization degree in the outermost ring becomes slightly higher.

\subsubsection{Effect of dust composition}
\label{subsubsec:results_dustcomp}

\begin{table}
	\caption{Dependence of the net polarization on the dust composition}              
	\label{tab:netpol_opac}      
	\centering
	\begin{tabular}{lcc}          
		\hline\hline                       
		Dust species & Net polarization [$\%$]\\
		\hline                                  
		Mixture & 0.042\\
		Silicate & 0.048\\
		Carbon & 0.25\\
		Water ice & 0.006\\
		\hline                                             
	\end{tabular}
	\tablefoot{The fractional abundances for the mixture are taken as 7\% silicate, 21\% carbon and 42\% water ice, so that the amount of vacuum is 30\%.}
\end{table}

\begin{figure*}
	\centering
	\centerline{
		\includegraphics[width=0.5\textwidth]{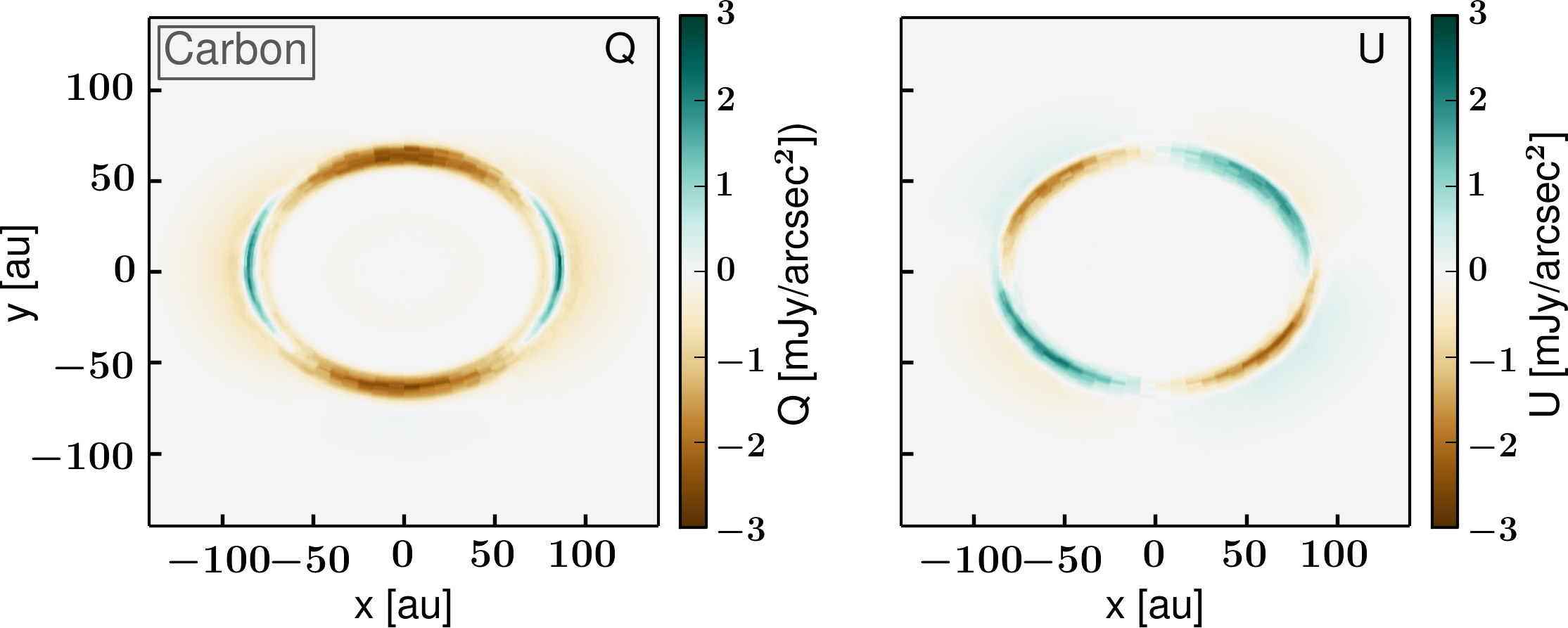}
		\includegraphics[width=0.5\textwidth]{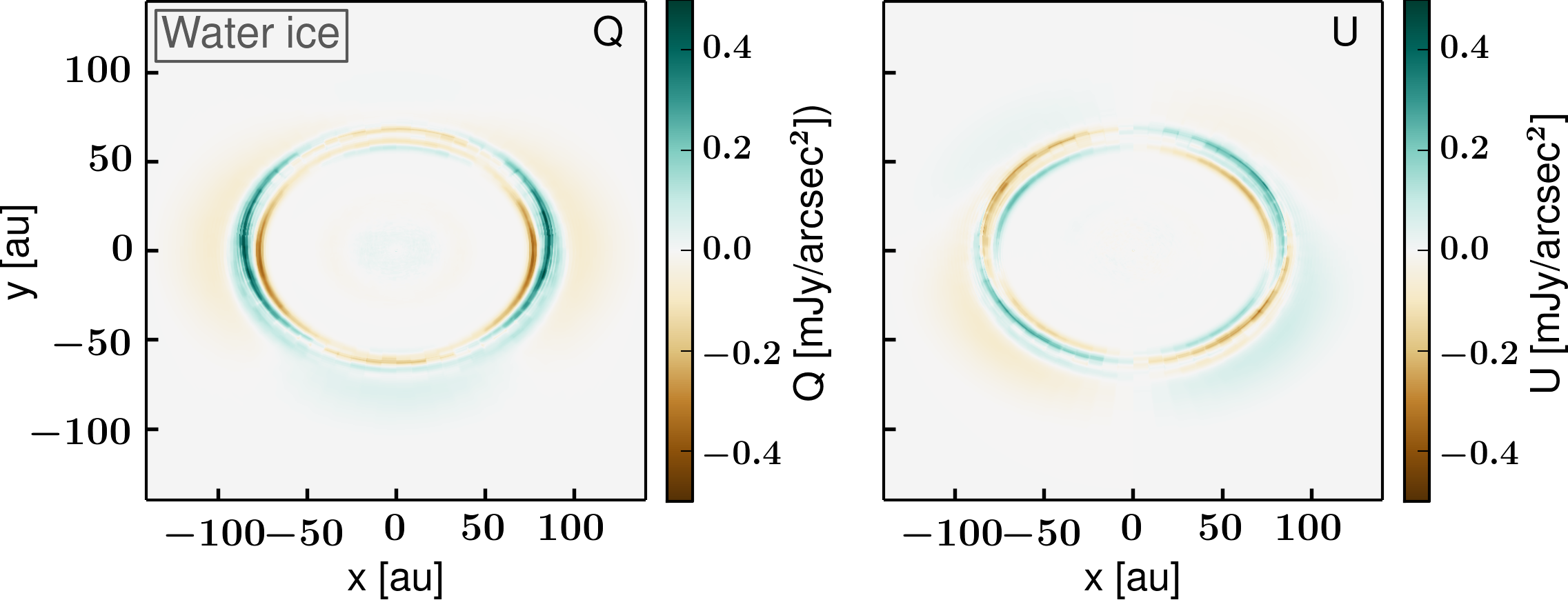}
	}
	\caption{Stokes $Q$ and Stokes $U$ maps for the models considering pure carbon (left) and pure water ice (right) opacities, respectively. Note the difference in the color map scaling.}
	\label{fig:qu_maps}
\end{figure*}

\begin{figure*}
	\centering
	\centerline{
		\includegraphics[width=0.33\textwidth]{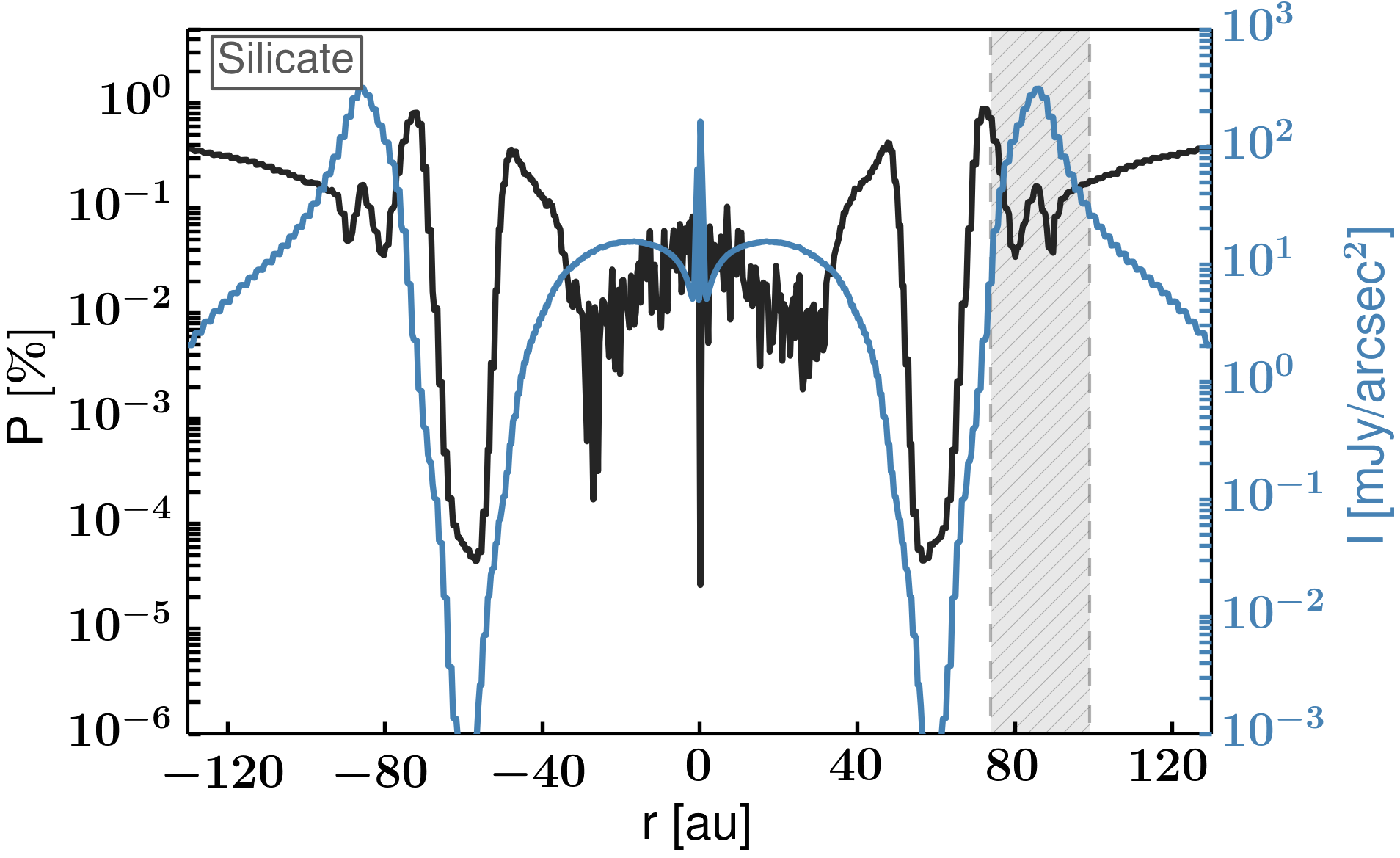}
		\includegraphics[width=0.33\textwidth]{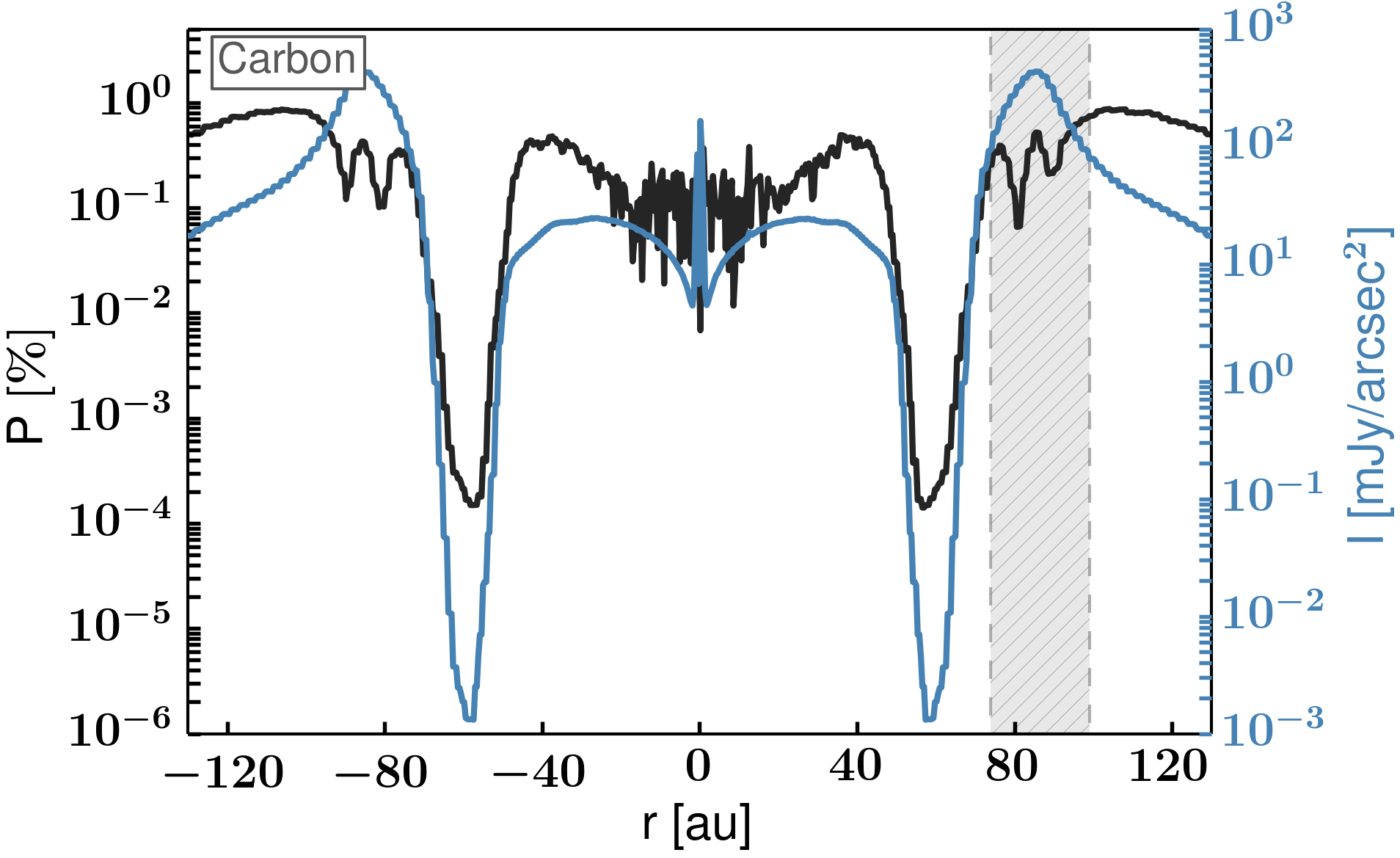}
		\includegraphics[width=0.33\textwidth]{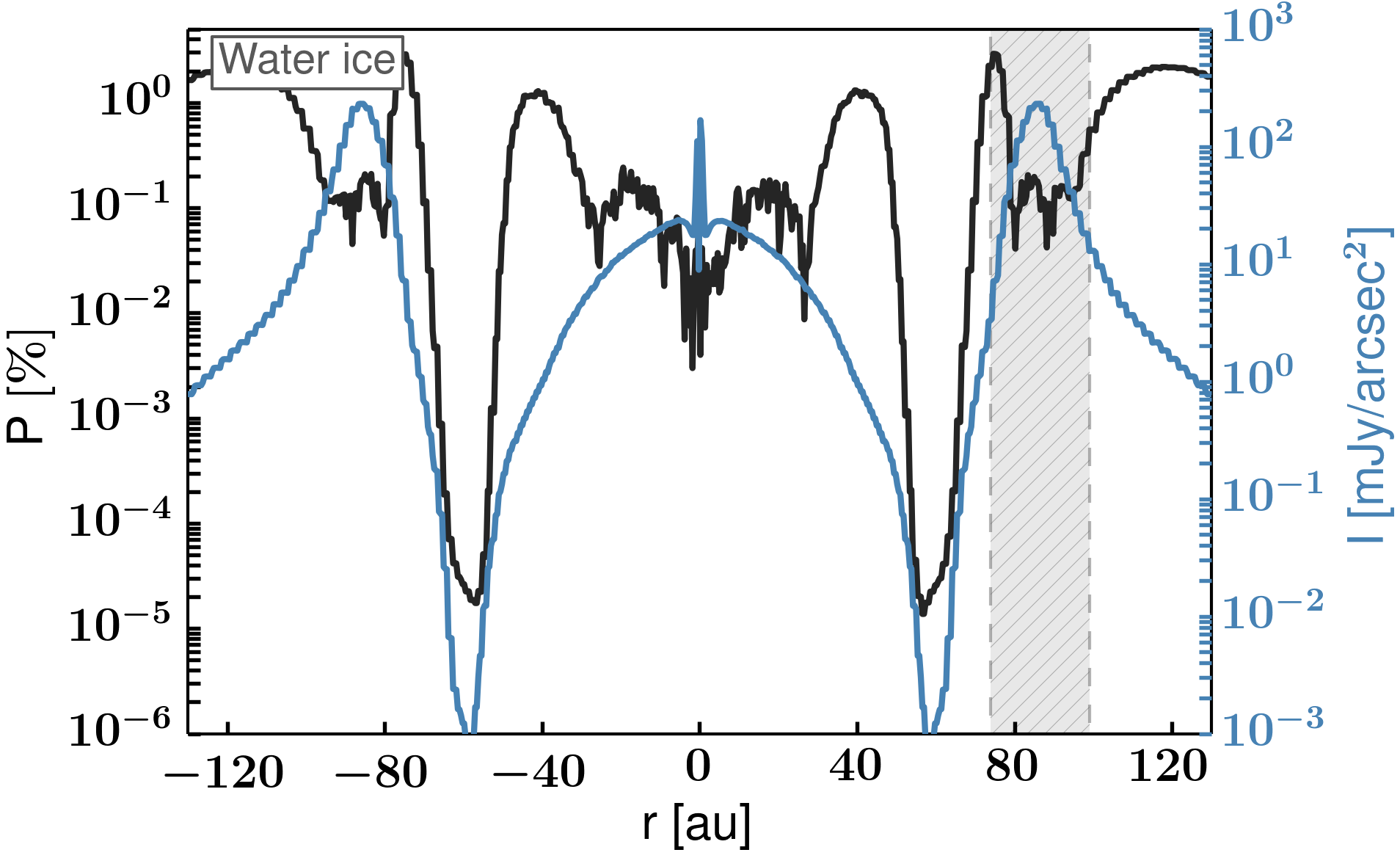}
	}
	\centerline{
		\includegraphics[width=0.33\textwidth]{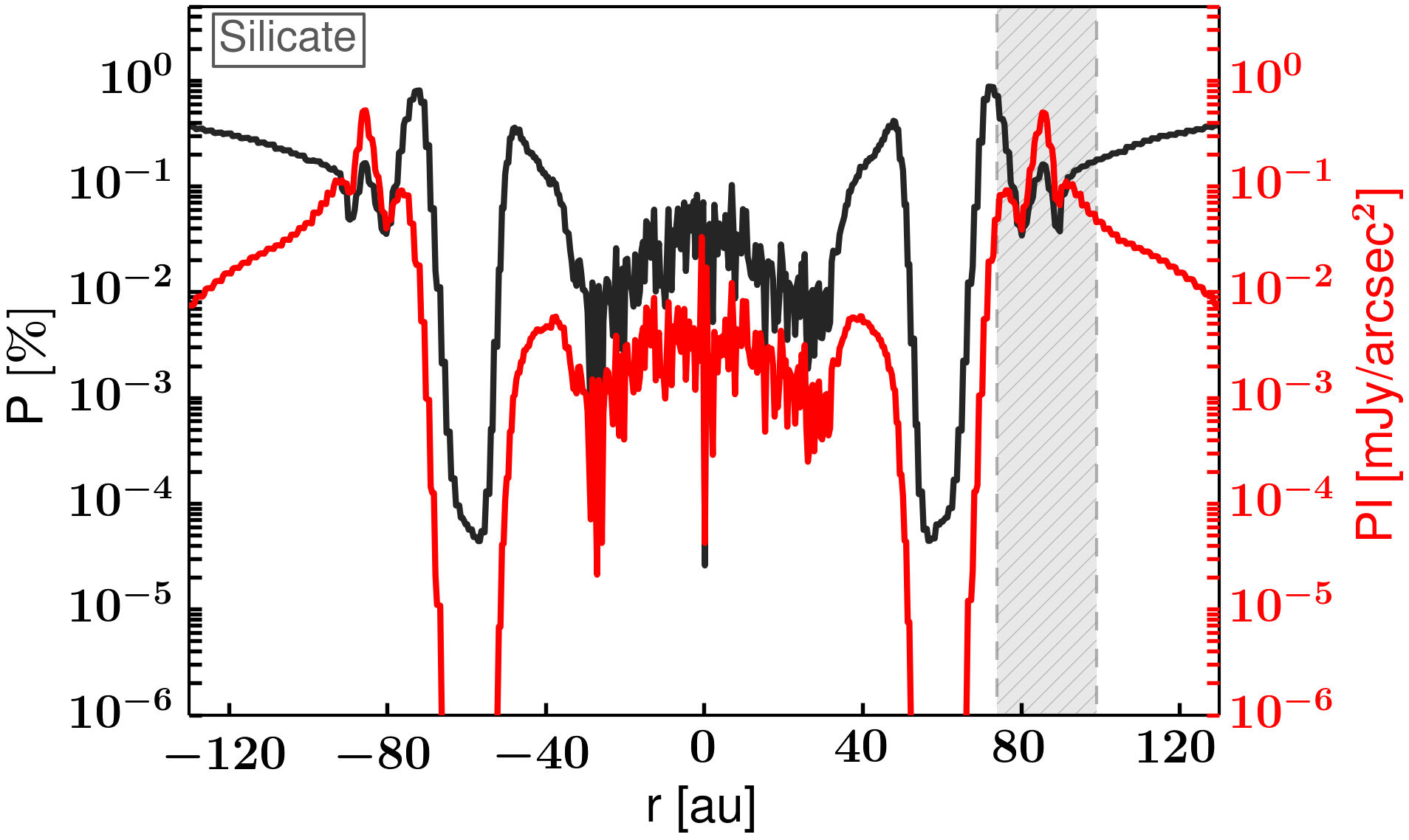}
		\includegraphics[width=0.33\textwidth]{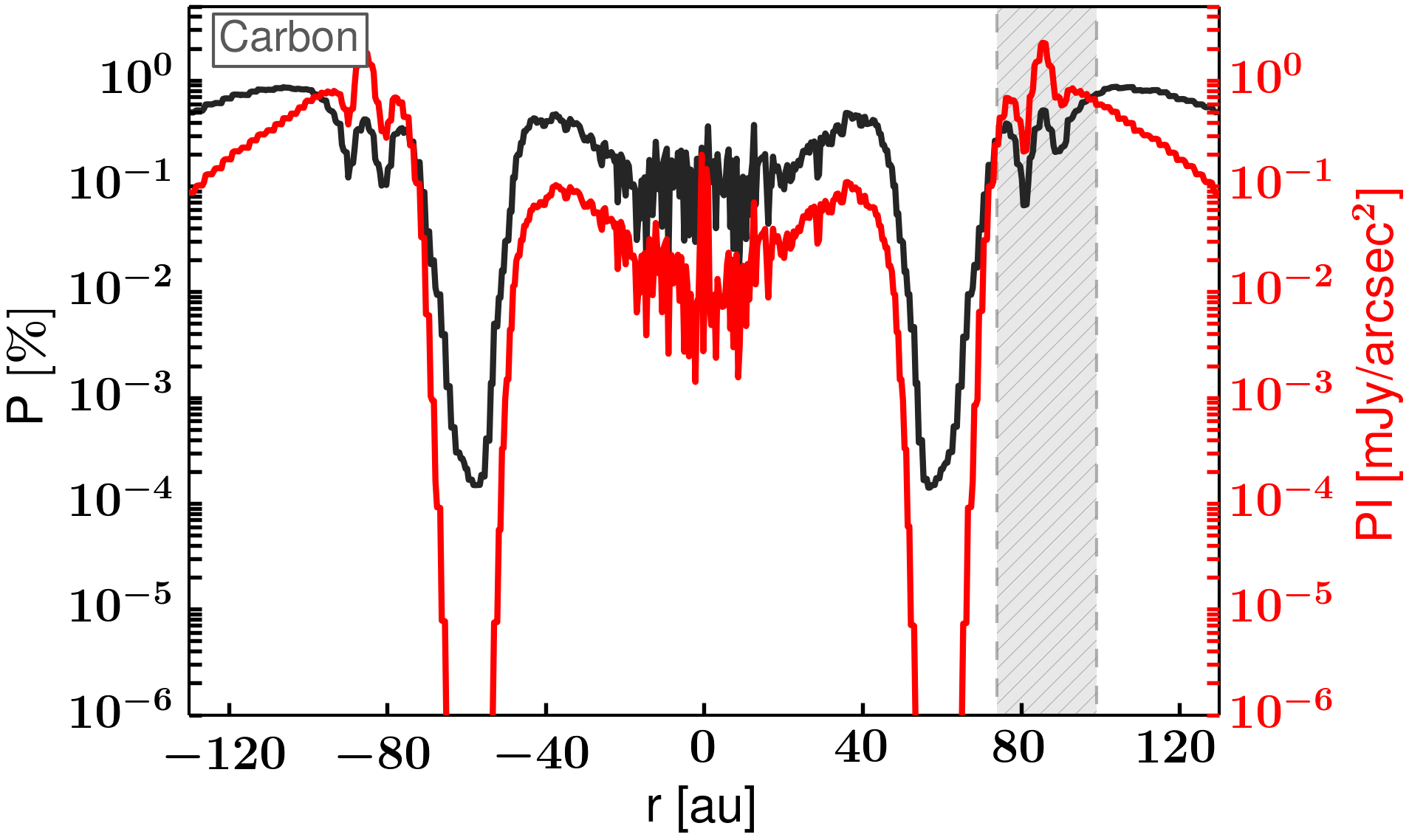}
		\includegraphics[width=0.33\textwidth]{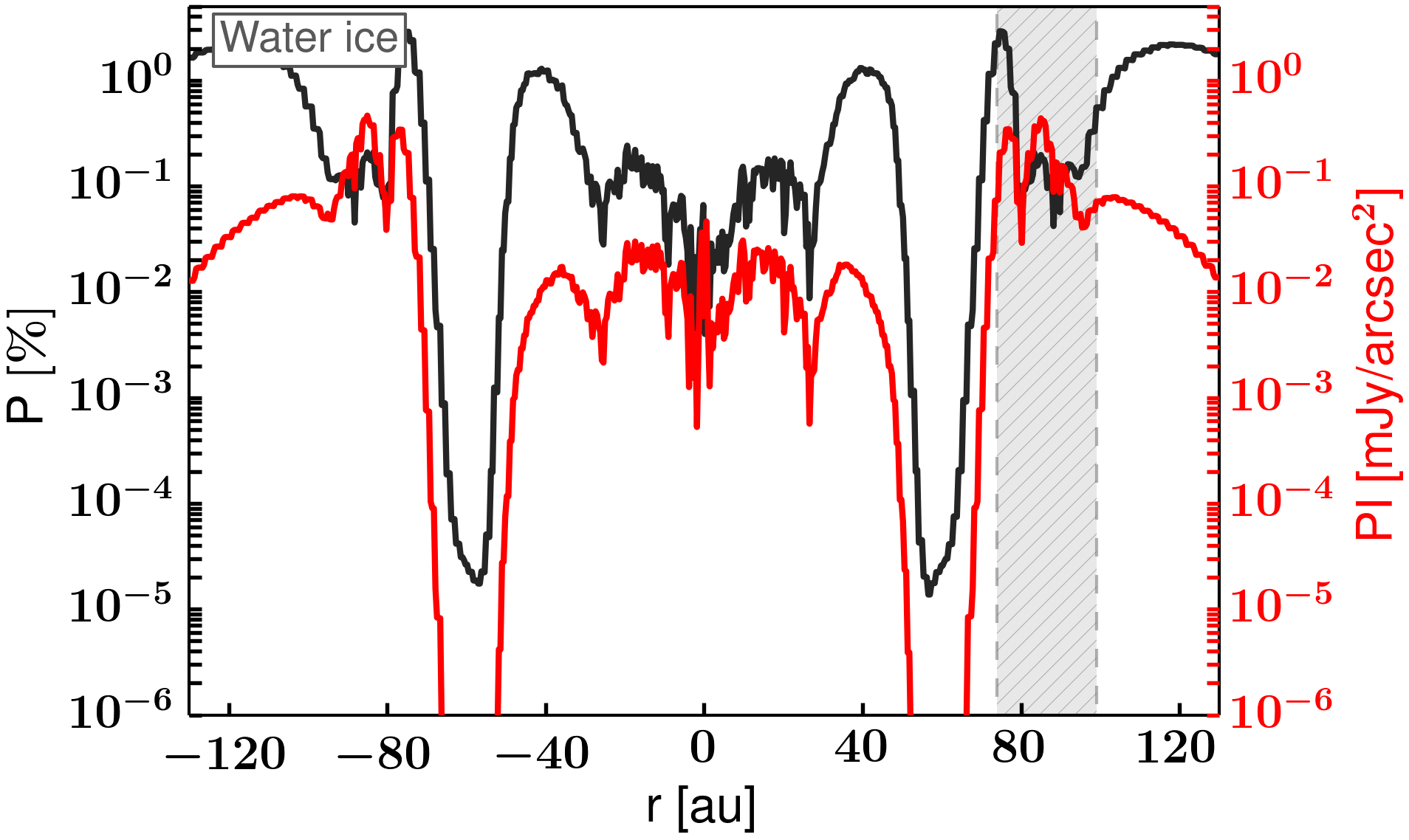}
	}
	\caption{\textit{Top row:} Radial cuts of the intensity (blue) and polarization degree (black) along the major axis. The gray crosshatched area visualizes the location of the highest intensity peak. \textit{Bottom row:} Radial cuts of the polarized intensity (red) and polarization degree (black) along the major axis. The gray shaded area again illustrates the highest polarized intensity peak.}
	\label{fig:intens_comp}
\end{figure*}

\begin{figure*}
	\centering
	\centerline{
		\includegraphics[width=0.33\textwidth]{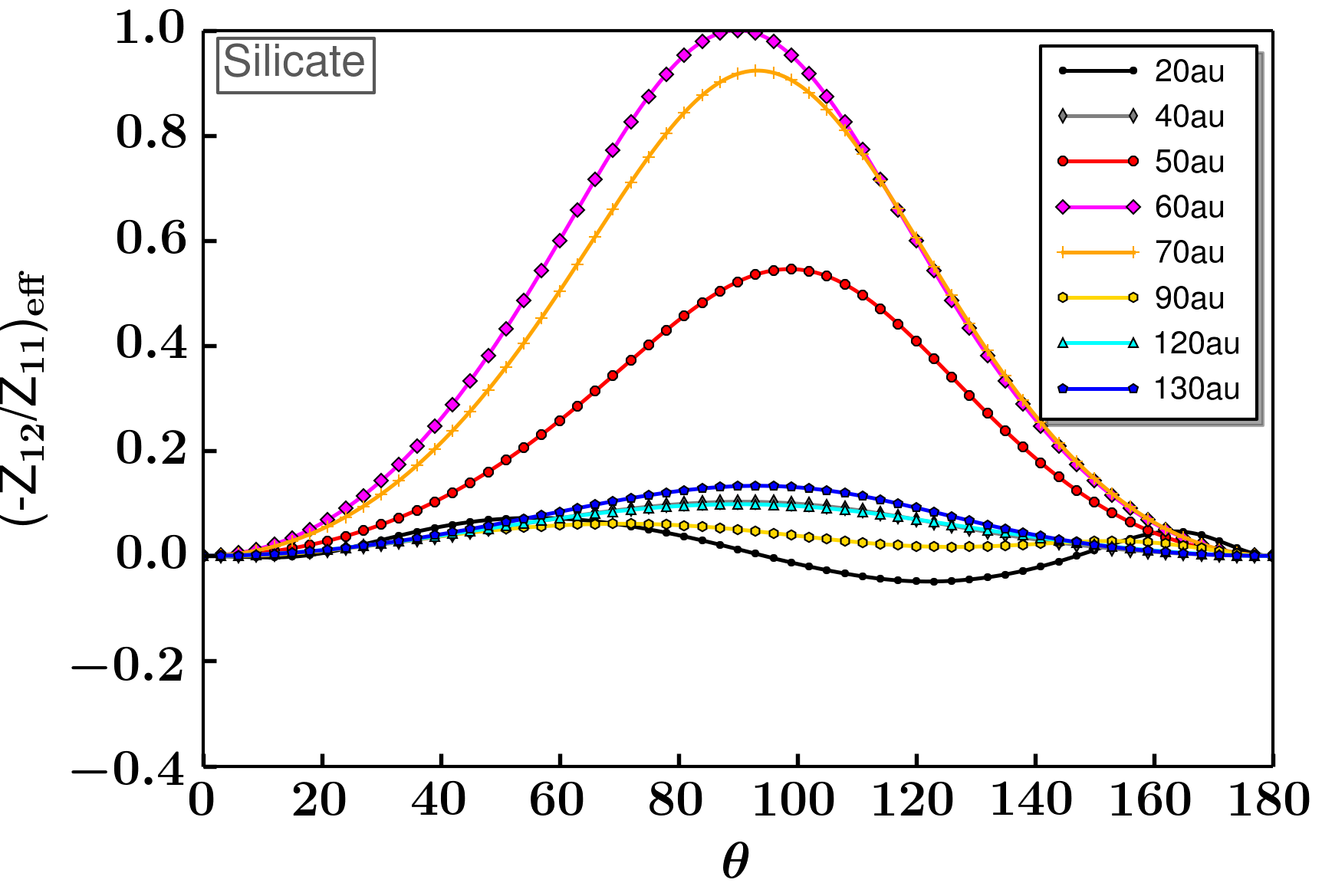}
		\includegraphics[width=0.33\textwidth]{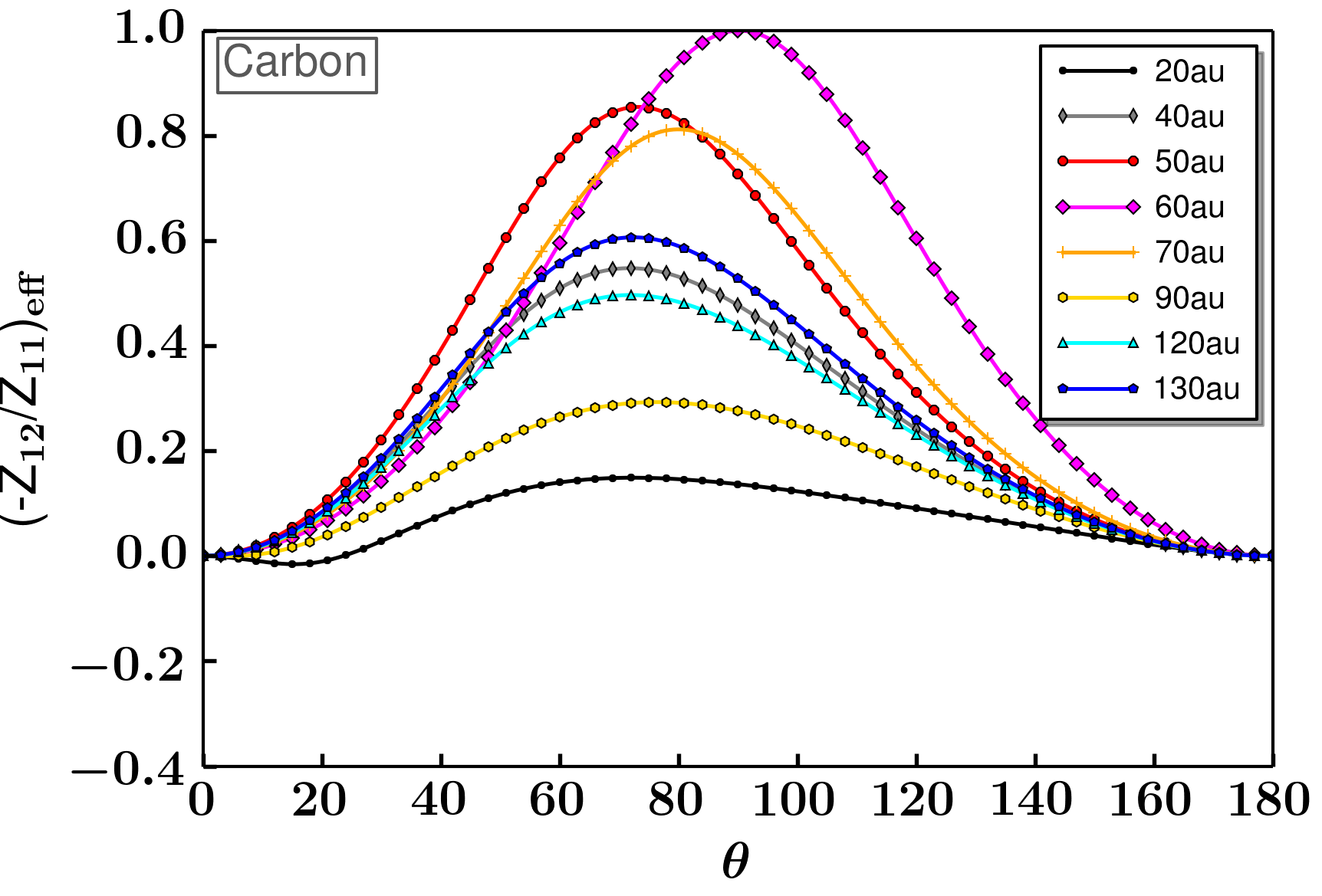}
		\includegraphics[width=0.33\textwidth]{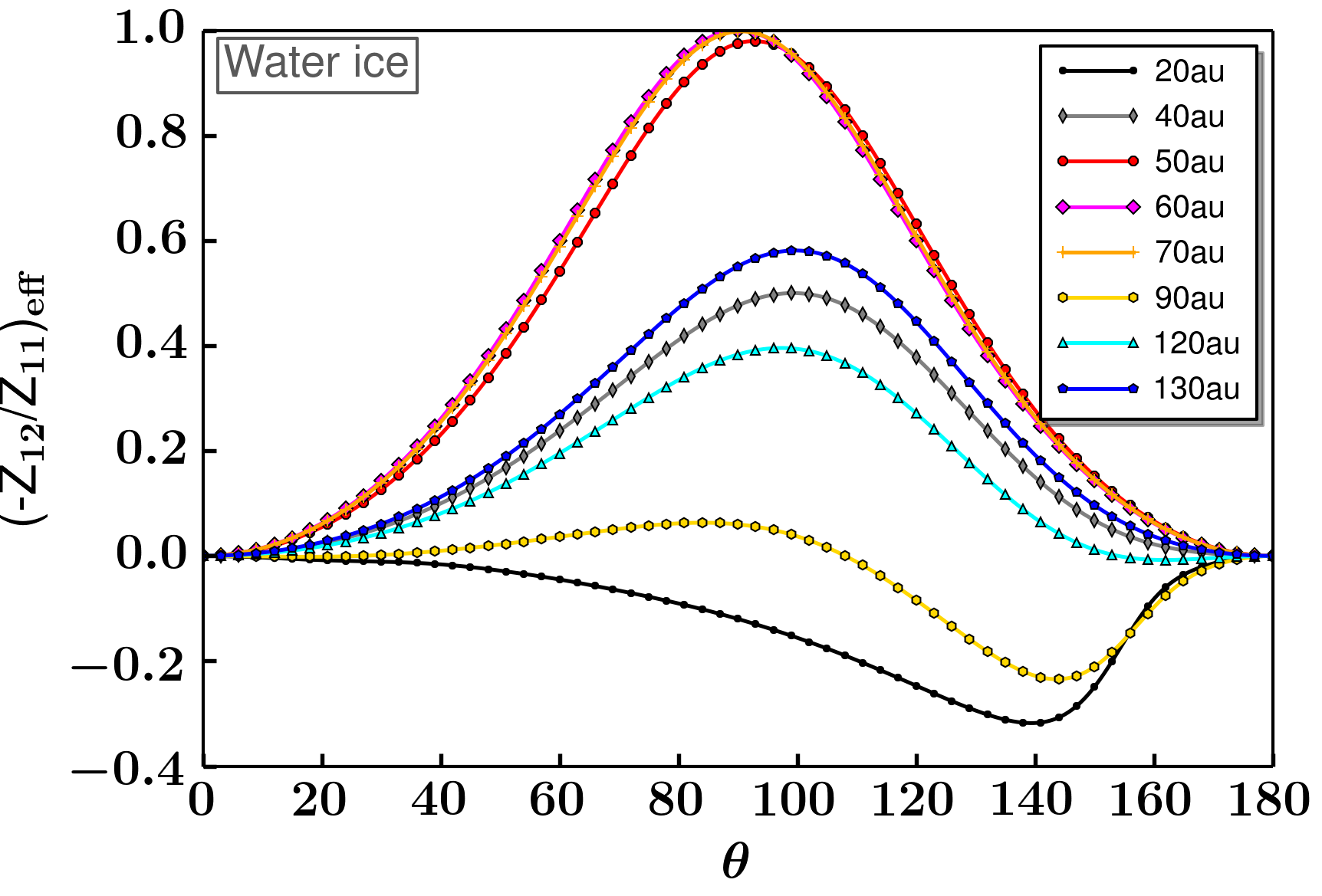}
	}
	\caption{Effective degree of polarization $(-Z_{12}(\theta)/Z_{11}(\theta))_{\mathrm{eff}}$ of different dust grain species in dependence on the scattering angle $\theta$. Several locations throughout the disk are chosen and highlighted in different colors.}
	\label{fig:z1211_comp}
\end{figure*}

We study the effect of the dust composition on the polarization by considering three `extreme' cases of pure silicate, pure carbon and pure water ice apart from our usual mixture (7\% silicate, 21\% carbon and 42\% water ice). We find that the dust composition has no effect on the overall polarization ring structure. In all cases, three polarization rings are produced at the described locations. This is a large advantage of dust-scattering as a method to constrain the grain sizes in protoplanetary disks. For a more quantitative study, the net polarization for the models at 40$^{\circ}$ inclination is listed in Table \ref{tab:netpol_opac}. At first sight it is remarkable that the values for the mixture and the pure silicate case are almost equal. This is not surprising, since the refractive index of the mixture is dominated by silicates due to its high material density, even though its volume fraction is small. Hence, the polarization curves plotted in Fig. \ref{fig:z1211_comp} for the silicate case and the ones for the dust mixture displayed in Fig. \ref{fig:ref_z1211} are comparable. As a consequence, for any mixture including silicates the polarization degree hardly depends on the dust composition. The situation changes for pure carbon with its high net polarization, and the opposite case of pure water ice, for which the lowest net polarization is found. This is explained by Fig. \ref{fig:qu_maps} illustrating the Stokes $Q$ and $U$ images. For carbon the negative $Q$ values (brown color) clearly exceed the amount of positive $Q$ (cyan color). Since the absolute value of $\Sigma Q$ is nearly two orders of magnitude higher than that for $\Sigma U$, the net polarization is determined by the Stokes $Q$ component and ends up being larger than zero. For water ice the summed up $Q$ value is dominating the $U$ component, however, it is still substantially lower than for carbon. These results are also supported by the polarized intensity profiles plotted in the bottom row of Fig. \ref{fig:intens_comp}. The polarized intensity peaks are highest for carbon and lowest for water ice, silicate is situated in-between. Furthermore, Fig. \ref{fig:z1211_comp} illustrates the effective degree of polarization. For carbon the polarization curves show the highest polarization degrees for all radii considered and no polarization reversal is detected between the relevant scattering angles of 50$^{\circ}$ and 130$^{\circ}$. Contrarily, the maxima for the polarization curves of pure water ice are lower and considerable negative polarization exists. Overall, these effects lead to the low net polarization of water ice compared to the mixture, the pure silicate and pure carbon case.\\

\begin{figure}
	\centering
	\includegraphics[width=0.5\textwidth]{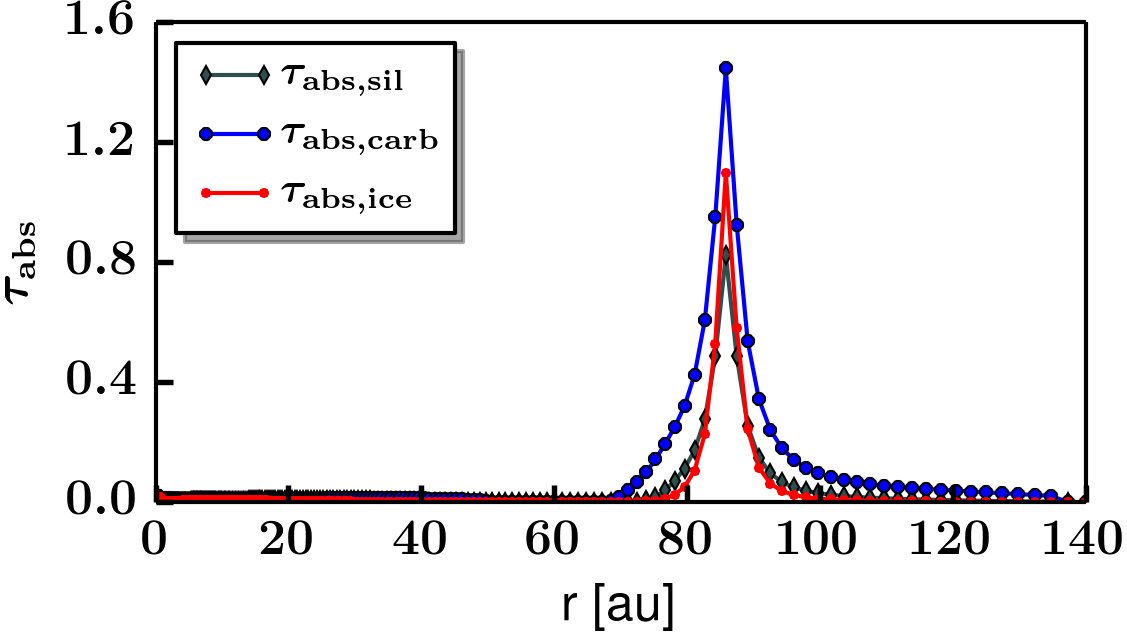}
	\caption{Optical depth $\tau_{\mathrm{abs}}$ for $\lambda = 0.87\,$mm in dependence on the radius for different dust species.}
	\label{fig:tau}
\end{figure} 

Apart from the net polarization there are also significant differences in the local polarization degree. We are again explaining the two limiting cases of pure carbon and water ice. Interestingly, the behavior of the net polarization calculations has turned. The polarization degree is plotted in Fig. \ref{fig:intens_comp} (black lines). It is basically determined by the flux gradients displayed in the radial intensity cuts in the top row of Fig. \ref{fig:ref_intens}, and by the effective polarization curves in Fig. \ref{fig:z1211_comp}. From the former one can understand the differences in the polarization degree of the innermost and outermost peak. The gradient of the intensity curve is much steeper for water ice leading to a higher polarization degree. Although the difference in the flux gradient is also slightly seen for the second polarization peak, it is noticeable that optical depth effects also come into play. In Fig. \ref{fig:tau} the optical depth in the vertical direction considering the vertically integrated dust density, $\tau_{\mathrm{abs}}(r) = \int_{a_{\mathrm{min}}}^{a_{\mathrm{max}}} \sigma_{\mathrm{d}}(a,r)\,\kappa_{\mathrm{abs}}(a)$, is plotted for $\lambda = 0.87\,$mm. At the position of the second ring ($\sim 70\,$au) the optical depth of carbon is larger than for water ice. Considering that we investigate inclined disks, which increases the optical depth, this means that the carbon model is marginally optically thick. Therefore, the corresponding radiation field is isotropic, so that we do not expect any polarization. Thus, the second peak in the polarization degree for water ice is also higher than the associated peak for the carbon case. 

\subsubsection{Effect of planet and its position}
\label{subsubsec:results_planetpos}

\begin{figure}
	\centering
	\includegraphics[width=0.5\textwidth]{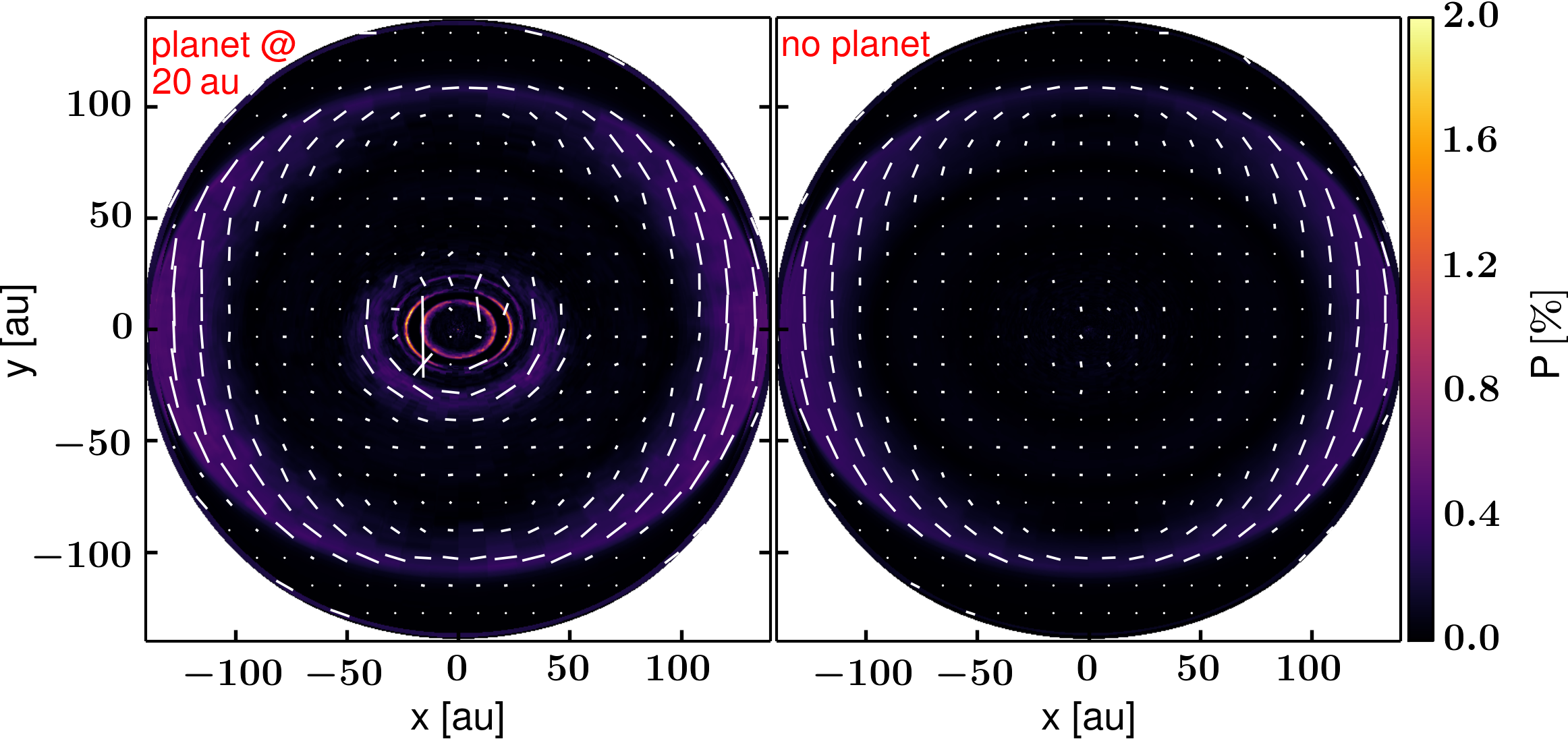}
	\caption{Polarization maps with overplotted polarization vectors for the disk model with the planet located at 20\,au and for a comparison model without any embedded planet, both maps are at $\lambda=0.87\,$mm. The maps are shown after 1\,Myr of dust evolution and for a disk inclination angle of 40$^{\circ}$.}
	\label{fig:pos_pol}
\end{figure}

As discussed in Sect. \ref{subsec:results_dust_distr} the presence of a planet triggers the trapping mechanism for mm-sized dust grains in the pressure bump outside of the planet orbit. Depending on whether the planet is embedded at 20\,au or 60\,au the gap width increases due to the radial density change and the pressure maximum is located at 30\,au and 90\,au, respectively. Figure \ref{fig:pos_pol} shows the polarization degree for the planet at 20\,au and a comparison model without any planet, for both cases at $\lambda = 0.87\,$mm. All other disk parameters are exactly the same as in the reference model. As mentioned before the two inner polarization rings are located just at the intensity gap edges. Hence, for the 20\,au case those rings are situated further inside with a narrower separation. Without any giant planet embedded in the disk only the outermost polarization ring remains. This is obvious since a gap and pressure bump in the dust distribution are missing as seen in Fig. \ref{fig:dustdensity_np}. There is only one crossing point between the horizontal lines and the maximum grain size layer. Therefore, polarization detection due to dust self-scattering can distinguish between transition disks hosting a gap and pressure bump on the one hand, and protoplanetary disks with a continuous surface density distribution on the other hand. We note that if the dust at the inside of the gap is depleted converting the gap into a complete cavity, the inner ring cannot be detected. Thus, in this case only the outer two rings are visible.

%------------------------------------------------------------------------------------

\section{Summary and conclusions}
\label{sec:conclusions}

We present a new technique to investigate the dust trapping scenario in transition disks hosting a pressure bump by means of mm-wave polarization of scattered thermal emission. More precisely, the continuum emission is polarized due to dust self-scattering of an anisotropic radiation field induced by disk inclination. For the mm-wave polarization to work the scattering grains must have grown to a maximum size of a few hundred microns. Additionally, further populations of either very small micron-sized or mm-/cm-sized grains need to be present in order to account for the large portion of unpolarized continuum emission. Our model predictions at different ALMA bands (bands 3, 6 and 7) are based on self-consistent dust growth models and radiative transfer calculations, which allows us to estimate the polarization degree in disks hosting a massive planet. The planet is considered to be located at 20\,au and 60\,au, respectively. Compared to previous studies a dust size distribution in radial direction is included in our model. Measuring the dust polarization degree is a direct and an unambiguous method to probe the location of large particles (a few hundred micron to mm sized particles, depending on the observing wavelength) when particle trapping occurs in the disk. We emphasize that the polarization technique presented can be applied to any disk with ring-like dust structures segregating the grain sizes, and is not limited to the gap opening scenario by planets. In this paper we focus on transition disks, because they are excellent candidates where planet formation may be ongoing. Thus, polarization observations are especially a unique tool to investigate the dust distribution when particle trapping is triggered by a planet embedded in the disk. The main findings of this paper are the following:

\begin{enumerate}
	\item The polarization pattern of a disk hosting a planetary gap after 1 Myr of dust evolution shows a characteristic three ring structure, where the two inner, narrow rings are located just at the gap edges. Additionally, there is a third polarization ring in the outer disk beyond 100\,au with a larger radial extension. Detecting such a distinctive ring structure with polarization observations may represent the radial size distribution of dust grains and hint to regions with a specific grain size corresponding to the wavelength.\\
		
	\item For an inclined disk there is an interplay between polarization originating from a flux gradient and from an inclination-induced anisotropy. The fraction of scattered radiation polarized due to disk inclination increases with the inclination angle. For intermediate inclined transition disks the polarization degree at $\lambda=3.1\,$mm is as high as $~2\,\%$, which is well above the detection limit of future ALMA polarization observations. A spatial resolution as high as 0\farcs2 is required to certainly resolve the ring structure in polarized intensity.\\
	
	\item The local degree of polarization is very sensitive to the maximum grain size at a certain location in the disk. Large cm and mm grains do not contribute to the polarization, which leads to a significant local reduction of the polarization degree, e.g. at the pressure bump region in our transition disk models. Hence, if the maximum grain size is larger than a few hundred microns, the polarization is not detectable, even if numerous small grains are present and contribute.\\
	
	\item For the face-on disks the polarization vectors are in azimuthal direction within the highest polarized intensity regions. For inclined disks the majority of the polarization vectors in the two inner rings are orientated along the minor axis. In the outer ring the gradient-induced polarization and, therefore, the azimuthal vector orientation still dominates.\\

	\item With increasing observing wavelengths the innermost and outermost polarization rings move inwards by a detectable distance, while the middle ring slightly moves radially outside. Hence, the positions of the two rings at the gap edges are shifted in opposite directions. For the outermost ring the moving distance is $\sim 20\,$au comparing the results at 0.87\,mm and 3.1\,mm. Furthermore, without any giant planet embedded in the disk, this third ring reflects the only polarized region, even though the polarization degree is quite low.\\  
	
	\item We find that the dust composition has no effect on the overall polarization ring structure. The presence of even a very small fraction of silicate in the dust mixture causes the local polarization degree and net polarization to be very similar to the case of pure silicate species. Silicate grains dominate the refractive index of the dust mixture, and the fractional abundances of carbonaceous material and water ice hardly affect the polarization pattern. A significant change of the local polarization degree with dust species can be led back to either optical depths effects or different flux gradients.

\end{enumerate}

We showed that polarization due to dust self-scattering is a powerful tool in order to constrain the grain size in protoplanetary disks independent of the spectral index. Nevertheless, there are also other polarization mechanisms such as alignment of dust grains along the magnetic field vectors. A keystone to investigate in future work is how to distinguish between these two mechanisms. Further multi-wave and spatially resolved polarization observations of protoplanetary disks are necessary. The wavelength and grain size dependence of polarization are also the keypoints from the modeling side. Due to the self-scattering mechanism we expect the polarization degree to significantly change with the observing wavelength, while it is thought to be approximately constant due to magnetically aligned grains in a toroidal magnetic field (\citealt{cho2007}).  

\begin{acknowledgements}
A.~P. is a member of the International Max Planck Research School for Astronomy and Cosmic Physics at Heidelberg University, IMPRS-HD, Germany. A.~K. acknowledges financial support from JSPS KAKENHI No. 15K17606. P.~P. is supported by Koninklijke Nederlandse Akademie van Wetenschappen (KNAW) professor prize to Ewine van Dishoeck.
\end{acknowledgements}

\bibliographystyle{aa}
\bibliography{aa_apohl}

\end{document}